\documentclass[twocolumn,tighten]{aastex62}

\def\kms{{km s$^{-1}$}}

\shorttitle{Galaxy Rotation Coherent with Environmental Dynamics}

\shortauthors{Lee et al.}
\def\simlt{\lower.5ex\hbox{$\; \buildrel < \over \sim \;$}}
\def\simgt{\lower.5ex\hbox{$\; \buildrel > \over \sim \;$}}

\begin{document}

\title{ Galaxy Rotation Coherent with the Motions of Neighbors: Discovery of Observational Evidence}

\author{Joon Hyeop Lee}
\email{jhl@kasi.re.kr}
\author{Mina Pak}
\author{Hye-Ran Lee}
\affil{Korea Astronomy and Space Science Institute, Daejeon 34055, Korea}
\affil{University of Science and Technology, Daejeon 34113, Korea}
\author{Hyunmi Song}
\affil{Korea Astronomy and Space Science Institute, Daejeon 34055, Korea}

\begin{abstract}
We present our discovery of observational evidence for the coherence between galaxy rotation and the average line-of-sight motion of neighbors. We use the Calar Alto Legacy Integral Field Area (CALIFA) survey data analyzed with the Python CALIFA {\footnotesize STARLIGHT} Synthesis Organizer (PyCASSO) platform, and the NASA-Sloan Atlas (NSA) catalog. After estimating the projected angular momentum vectors of 445 CALIFA galaxies, we build composite maps of their neighbor galaxies on the parameter space of line-of-sight velocity versus projected distance. The composite radial profiles of the luminosity-weighted mean velocity of neighbors show striking evidence for dynamical coherence between the rotational direction of the CALIFA galaxies and the average moving direction of their neighbor galaxies. The signal of such dynamical coherence is significant for the neighbors within 800 kpc distance from the CALIFA galaxies, for which the luminosity-weighted mean velocity is as large as $61.7\pm17.6$ {\kms} ($3.5\sigma$ significance to bootstrap uncertainty) when the angular momentum is measured at $R_e<R\le 2R_e$ of each CALIFA galaxy. In the comparison of the subsamples, we find that faint, blue or kinematically misaligned galaxies show stronger coherence with neighbor motions than bright, red or kinematically well-aligned galaxies do. Our results indicate that (1) the rotation of a galaxy (particularly at its outskirt) is significantly influenced by interactions with its neighbors up to 800 kpc, (2) the coherence is particularly strong between faint galaxies and bright neighbors, and (3) galaxy interactions often cause internal kinematic misalignment or possibly even kinematically distinct cores.
\end{abstract}


\section{INTRODUCTION}\label{intro}

The internal kinematics of a galaxy is an important factor to understand the formation history of the galaxy. Over the classical understanding of \emph{pressure-supported} early-type galaxies and \emph{rotation-supported} late-type galaxies, it has been revealed that even early-type galaxies usually rotate and they can be divided into slow and fast rotators, which has risen as a quite new point of view on galaxy classification \citep{cap06,ems07}. Many subsequent studies enhanced such an idea by investigating more details of slow and fast rotators, and even proposed that the famous Hubble tuning fork needs to be revised, not only because it tends to mislead our understanding of early-type galaxies by ignoring the large variety of fast rotators, but also because the new \emph{kinematic morphology} shows a tighter relationship with environmental density than classical photometric morphology does \citep{cap11,ems11,kra11,fog14,cor16,bro17,gre17,van17,fos18,gra18,ron18,sme18}. Like these, with the advent of integral field spectroscopy (IFS), our understanding of galaxy kinematics has been being improved very rapidly for a recent decade.

One of the key issues about galaxy kinematics is the effect of environments on it.
It is well known that environmental effects, from pair interactions/mergers to large-scale mechanisms, play important roles in galaxy evolution, through many observational studies \citep[e.g.,][]{dre80,kau04,bal06,bla07,pog08,lee10,lee14,pop11,sco13,lee16,pak16,fu18} and numerical predictions \citep[e.g.,][]{jun14,gen16,mar18}.
Thus, it is natural to expect that environmental effects play some crucial roles in galaxy kinematics, as well.
In recent observational studies, for example, it was argued that an equal-mass polar merger may result in galaxy rotation around major photometric axis \citep[prolate rotation;][]{tsa17,kra18,wea18}.
Furthermore, interactions or mergers between galaxies are thought to be also responsible for the misalignment between the morphological shape and rotational direction in a galaxy \citep[morpho-kinematic misalignment;][]{bar15,oh16}.
In addition to the morpho-kinematic misalignment, the angular momentum in a galaxy may not be perfectly aligned along radius (i.e., internal kinematic misalignment), the extreme cases of which may be classified as kinematically distinct cores \citep[KDCs;][]{ems07,ems14,kra15}. In recent numerical studies, it was shown that KDCs can result from major or minor merging events \citep{boi11,tay18}.
Various observational studies argued that galaxy environments may have played an important role in determining the internal kinematics of galaxies either in a small scale \citep{lee18b} or in a large scale \citep{hou13,kim18}, which is also supported by hydrodynamic simulations \citep{naa14,lag17,lag18a,lag18b,lee18a}.

Those recent accomplishments about the environmental effects on galaxy kinematics lead us to a simple and basic question: if galaxy kinematics is largely influenced or even determined by environments, then can we find any coherence between the rotational direction of a galaxy and the motions of its neighbor galaxies? In other words, is there any observational evidence that galaxy interactions directly change the rotational direction of a galaxy?
If the answer is `yes', then how significantly is galaxy rotation affected by environments and how is such dynamical coherence related with other properties of galaxies in detail? 
Recently, \citet{lee18a} showed that tidal perturbations as well as merging events may significantly change galaxy spin vectors in their hydrodynamic simulations. In observational studies, however, those questions have never been clearly answered yet, despite their importance to clarify the origin of galaxy rotation.

Thus, in this paper, we examine the coherence between the rotation of a galaxy and the line-of-sight motions of its neighbors, using publicly available IFS and spectroscopic surveys.
This paper is outlined as follows. Section~2 describes the archival data used in this paper. Section~3 shows how we utilize the data in order to find the signal of the dynamical coherence. The results are presented in Section~4. What our results imply is discussed in Section~5, and the paper is concluded in Section~6.
Throughout this paper, we adopt the cosmological parameters: $h=0.7$, $\Omega_{\Lambda}=0.7$, and $\Omega_{M}=0.3$.

\section{DATA AND METHODS}\label{data}

\begin{figure}[t]
\centering
\plotone{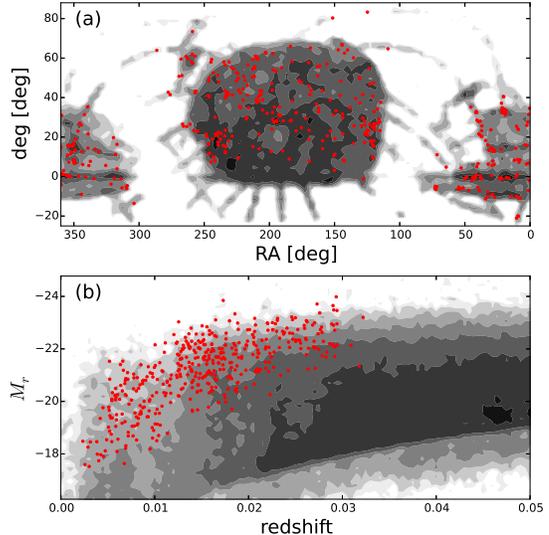}
\caption{(a) Spatial distribution and (b) redshifts and $r$-band absolute magnitudes of the CALIFA galaxies (red dots). The background contours show the log-scale number density of the NSA galaxies.\label{spdist}}
\end{figure}

\begin{figure}[t]
\centering
\plotone{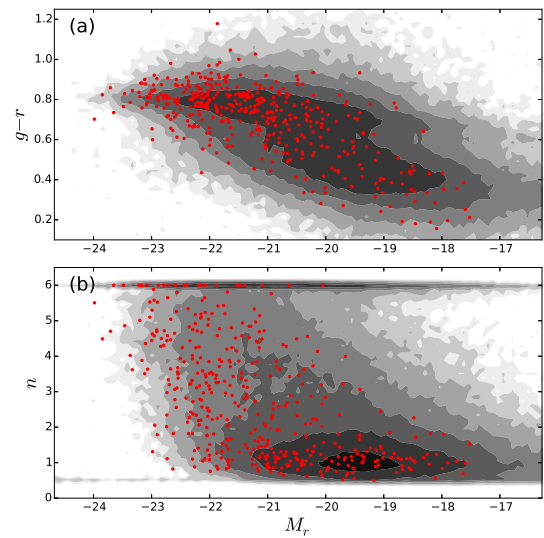}
\caption{(a) Color-magnitude diagram and (b) S{\'e}rsic index distribution as a function of $r$-band absolute magnitude of the CALIFA galaxies (red dots) and the NSA galaxies (background contours).\label{cmr}}
\end{figure}

To investigate the relationship between galaxy rotation and neighbor motions, two sorts of information are necessary: the internal kinematics of a sufficient number of galaxies and the motions of their neighbor galaxies. To obtain the former, a large IFS survey of galaxies is required, while the latter can be acquired from a huge spectroscopic survey that includes the IFS targets and their neighbors.
Fortunately, today both kinds of data sets can be conveniently acquired owing to the publicly-released data of various surveys.
Among several available IFS surveys, the Calar Alto Legacy Integral Field Area Survey \citep[CALIFA;][]{san12,san16,wal14} is particularly suitable for this purpose, because its spatial coverage for each target is very large (frequently larger than twice of half-light radius). Such a large coverage makes it possible to investigate the radial variation of internal kinematics in each target. Since the target selection of the CALIFA is based on the Sloan Digital Sky Survey \citep[SDSS;][]{yor00}, the information of the targets' neighbor galaxies can be retrieved from the SDSS and additional spectroscopic surveys.

\subsection{CALIFA and PyCASSO Database}

To acquire the angular momentum vectors of a sufficient number of galaxies, we use the CALIFA survey data\footnote{http://califa.caha.es/}. The CALIFA sample consists of $\sim600$ galaxies at low redshifts, which have been selected from the photometric catalog of the SDSS as a sample limited in apparent isophotal diameter. The targets were observed using the PMAS/PPAK integral field spectrophotometer, mounted on the Calar Alto 3.5 m telescope. The spectra cover the wavelength range of $3700-7000$ {\AA}.
The most outstanding strength of the CALIFA survey is the extremely wide field-of-view ($> 1$ arcmin$^2$) with a high filling factor in a single pointing ($65\%$). This provides great advantage for the studies of low-redshift galaxies with large angular sizes, because CALIFA observations cover a considerably large area in each target (mostly much larger than half-light radius).
For more details of the CALIFA survey, refer to \citet{san12,san16}.

In this paper, the actual analysis is based on the PyCASSO database\footnote{http://pycasso.ufsc.br or http://pycasso.iaa.es/} \citep{dea17}, which is a data set value-added by analyzing the CALIFA data with the Python CALIFA {\footnotesize STARLIGHT} Synthesis Organizer platform \citep{cid05,cid13}. The sample of the PyCASSO database consists of 445 galaxies in the CALIFA Data Release 3 sample \citep{san16}. Those sample galaxies were observed with both V500 and V1200 gratings and their combination (called \emph{COMBO} cubes), all of which are necessary to reduce the vignetting and to fill the whole field of view through a dithering scheme. More details are described in \citet{dea17}.
Although the sample size of the PyCASSO database is smaller than that of the full CALIFA data set, its well-produced final products are very useful for quick and reliable investigation.

The PyCASSO database provides various parameter maps of 445 CALIFA galaxies, which includes  stellar mass surface density, line-of-sight velocity, and some stellar population indicators such as mean age and metallicity. In this paper, stellar mass ($M_*$) surface density and line-of-sight velocity ($v_*$) distributions are used in order to estimate the angular momentum vectors of those galaxies.
When the signal-to-noise (S/N) is $20 - 30$, the typical uncertainties for these quantities are: 0.09 for $\log M_*$ and 19 km s$^{-1}$ for $v_*$, while they are as small as 0.04 for $\log M_*$ and 9 km s$^{-1}$ for $v_*$ when the S/N is $40 - 50$ \citep{cid14,dea17}.
Figure~\ref{spdist} shows the distributions of the 445 CALIFA galaxies in the sky and in the absolute magnitude versus redshift diagram. The CALIFA galaxies are evenly distributed over the spatial coverage of the SDSS, and they have very low redshifts ($z\lesssim0.03$).
Figure~\ref{cmr} presents the color and S{\'e}rsic index distributions as a function of absolute magnitude. The CALIFA galaxies distribute over both red sequence and blue cloud, and their morphological types are not significantly biased to any of early-type or late-type. All of the CALIFA galaxies are brighter than $M_r=-17$ mag.

\begin{figure}[t]
\centering
\plotone{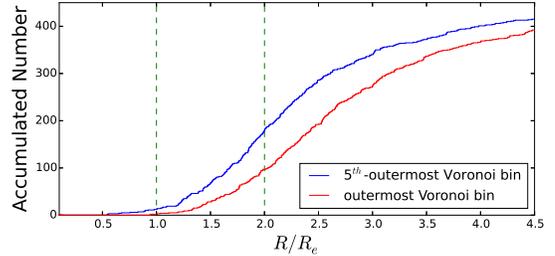}
\caption{Statistics of the CALIFA spatial coverage in individual galaxies. The red line accumulates the number of CALIFA galaxies as a function of the radial distance (normalized by half-light radius, $R_e$) to the outermost Voronoi bin in each galaxy, while the blue line does it to the fifth-outermost Voronoi bin.\label{cover}}
\end{figure}

Figure~\ref{cover} shows how large the spatial coverage in the CALIFA observation for each target is. Among the 445 CALIFA galaxies, only two galaxies have the spatial coverage slightly smaller than half-light radius ($R_e$), while the covered areas for 348 galaxies ($78\%$) are larger than $2R_e$. 433 galaxies ($97\%$) have at least five Voronoi bins at $R>R_e$, which means that the outskirt ($R_e<R\le 2R_e$) angular momenta can be measured for a considerable fraction of the CALIFA sample.

\begin{figure}[t]
\centering
\plotone{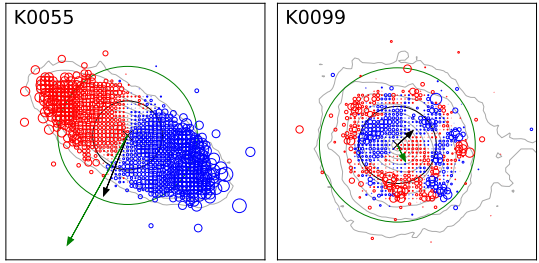}
\caption{Two examples of galaxy angular momentum measurement. The sizes of the red (representing redshift) and blue (representing blueshift) circles are proportional to the radial velocity relative to each galactic center. The black circles show the half-light radii, while the green circles correspond to twice of the half-light radii. The black arrows represent the projected angular momentum vectors estimated using the Voronoi bins at $R{\le}R_e$, while the green arrows are those at $R_e<R\le 2R_e$. The gray contours show the galaxy surface brightness distributions.\label{angmom}}
\end{figure}

\begin{figure}[t]
\centering
\plotone{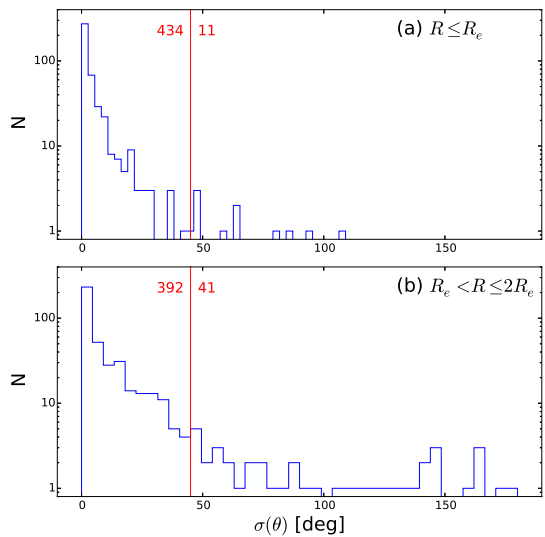}
\caption{Uncertainty distribution of the position angles of angular momentum vectors for (a) $R\le R_e$ and (b) $R_e<R\le 2R_e$. The vertical lines denotes 45$^{\circ}$ uncertainty. The numbers of galaxies with position angle uncertainties smaller and larger than 45$^{\circ}$ are denoted at the left and right sides of the line, respectively.\label{angerr}}
\end{figure}

\begin{figure}[t]
\centering
\plotone{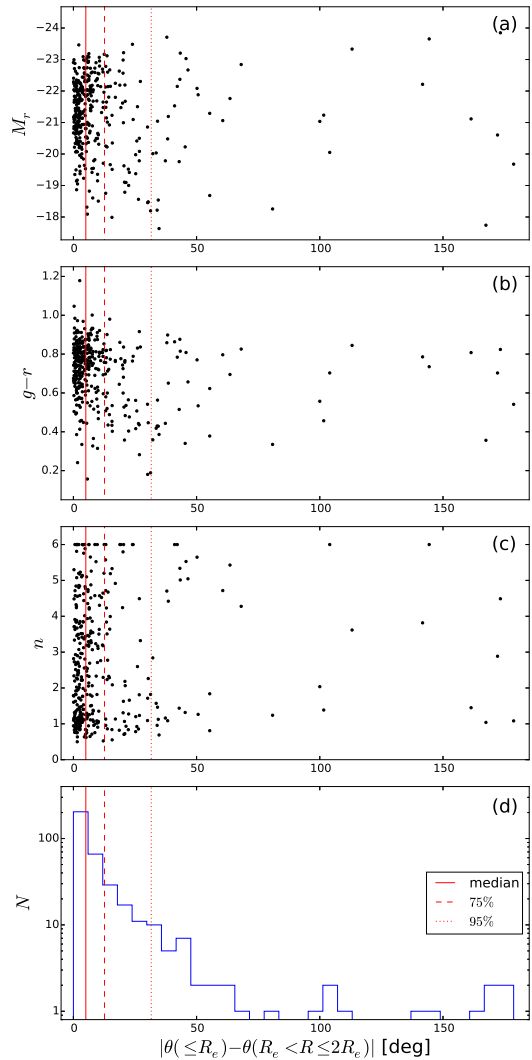}
\caption{Internal angular momentum misalignment of individual CALIFA galaxies. (a) Absolute magnitude in the $r$ band, (b) $g-r$ color, (c) S{\'e}rsic index, and (d) number distribution as a function of misalignment angle. The median value among the misalignment angle distribution is $5.0^{\circ}$. \label{misalign}}
\end{figure}

\subsubsection{Projected Angular Momentum}

Using the PyCASSO database, we estimated the projected angular momentum vectors ($\mathbf{L} = \mathbf{r} \times \mathbf{p} = \mathbf{r} \times m\mathbf{v}$) of the CALIFA galaxies. For simplification, we supposed that the mass and velocity distribution in a given Voronoi bin are perfectly represented by a point mass located at the center of the bin, which has the total mass and the mean stellar line-of-sight velocity measured in the bin.

Figure~\ref{angmom} shows two examples of the projected angular momentum estimation.
In each galaxy, we derived two projected angular momentum vectors, which were estimated from the central area ($R{\le}R_e$) and the outskirt area ($R_e<R\le 2R_e$), respectively.
The estimation of an angular momentum vector is simple and direct. However, the estimated position angle has uncertainty, which may be particularly large for some slow rotators. Thus, we estimated the statistical uncertainty of the position angle, through repetitive random sampling with replacement of spaxels. As presented in Figure~\ref{angerr}, a majority of the CALIFA galaxies show position angle uncertainties smaller than 45$^{\circ}$: 434/445 ($98\%$) for $R{\le}R_e$ and 392/433 ($91\%$) for $R_e<R\le 2R_e$ (12 of 445 CALIFA galaxies have less than 5 Voronoi bins at $R>R_e$).  The galaxies with position angle uncertainties larger than 45$^{\circ}$ are excluded in our analysis. Those galaxies are not guaranteed to have position angle errors smaller than 90$^{\circ}$ in $2\sigma$ confidence level, which means that they may falsely add opposite signals to our results by possibility higher than $5\%$.

In Figure~\ref{angmom}, one example (K0055) shows quite good alignment between the two vectors, whereas the other (K0099) has large misalignment between the two vectors.\footnote{In many studies, `kinematic misalignment' often indicates the difference between morphological position angle and kinematic position angle of a galaxy (often denoted by $\Psi$). However, in this paper, the `misalignment' indicates the difference between central kinematic position angle ($R\le R_e$) and outskirt kinematic position angle ($R_e<R\le 2R_e$) in a galaxy; a large misalignment angle corresponds to a kinematically distinct core.}
Figure~\ref{misalign} presents the distribution of such internal misalignment of the CALIFA galaxies, which is compared with the distributions of absolute magnitude, color and S{\'e}rsic index.
Although a half of the CALIFA galaxies have very small misalignment angles ($<5^{\circ}$), about $5\%$ of them have misalignment angles larger than $30^{\circ}$. However, such internal misalignment of angular momentum does not appear to be correlated with luminosity, color or morphology of those galaxies in Figure~\ref{misalign}. Largely-misaligned galaxies show a notable bimodality in the S{\'e}rsic index distribution (that is, the galaxies with intermediate $n$ tend to be well aligned), but it is clear that the misalignment is not a monotonic function of any parameters examined here.

\subsection{NASA-Sloan Atlas}\label{morcla}

To estimate the line-of-sight motions of the neighbor galaxies around the CALIFA galaxies, we use the NASA-Sloan Atlas (NSA) catalog\footnote{http://www.nsatlas.org}, which was created by Michael Blanton, based on the SDSS, NASA Extragalactic Database (NED)\footnote{https://ned.ipac.caltech.edu/, which is operated by the Jet Propulsion Laboratory, California Institute of Technology, under contract with the National Aeronautics and Space Administration.}, Six-degree Field Galaxy Redshift Survey \citep[6dFGS;][]{jon09}, Two-degree Field Galaxy Redshift Survey \citep[2dFGRS;][]{col01}, CfA Redshift Survey \citep[ZCAT;][]{huc83}, Arecibo Legacy Fast ALFA Survey \citep[ALFALFA;][]{gio05} and the Galaxy Evolution Explorer \citep[GALEX;][]{mar03} survey data.
Among the numerous parameters provided by the NSA catalog, we use the information of right ascension, declination, redshift, S{\'e}rsic index and absolute magnitudes in the $g$ and $r$ bands.
Line-of-sight velocity offset ($\Delta v$) is calculated from redshifts of a neighbor galaxy ($z_{\textrm{\tiny nei}}$) and a CALIFA galaxy ($z_{\textrm{\tiny CAL}}$), based on the assumptions that (i) the peculiar motion of a neighbor is non-relativistic; (ii) a CALIFA galaxy does not have peculiar motion (follows the Hubble expansion only); and (iii) gravitational redshift is negligible. The conversion equation is as follows:
\begin{equation}
 \Delta v = \frac{z_{\textrm{\tiny nei}} - z_{\textrm{\tiny CAL}}}{1+z_{\textrm{\tiny CAL}}} \cdot c,
\end{equation}
where $c$ is the speed of light.

\begin{figure*}[t]
\centering
\includegraphics[width=0.95\textwidth]{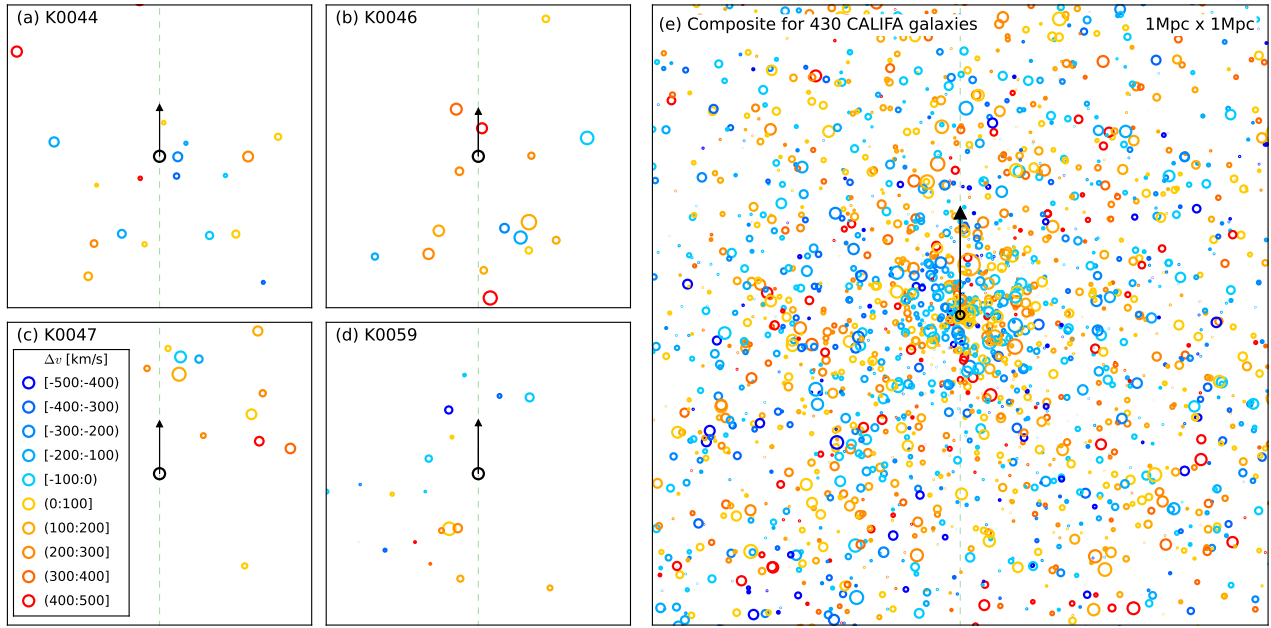}
\caption{Process to build a composite map of neighbor distribution. (a) - (d) Examples of individual CALIFA galaxies (black circles) and their neighbors with $\Delta v{\le}500$ {\kms} in 1 Mpc $\times$ 1 Mpc areas. The color and size of each circle represent the line-of-sight velocity and luminosity of each neighbor galaxy relative to a given CALIFA galaxy, respectively. The spatial distribution of the neighbor galaxies is aligned so that the projected angular momentum vector of the CALIFA galaxy (black arrow) is upward. (e) Composite map of neighbor galaxy distribution for our whole sample of the CALIFA galaxies. \label{stack}}
\end{figure*}

\begin{figure}[t]
\centering
\plotone{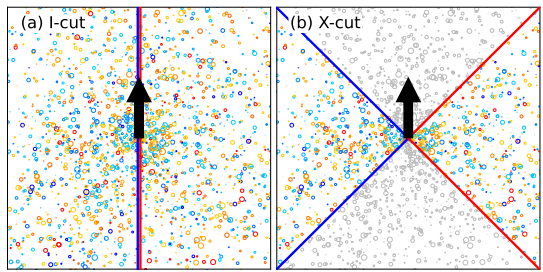}
\caption{Definition of areas to estimate neighbor motions. (a) {\it I-cut} simply divides the entire area into two sub-areas using a vertical line. (b) {\it X-cut} excludes the neighbors in the domains of $-45^{\circ}<\theta<45^{\circ}$ and $135^{\circ}<\theta<225^{\circ}$, where $\theta$ is the position angle from the angular momentum vector direction. \label{cuts}}
\end{figure}

\section{ANALYSIS}\label{anal}

To explore the dynamical correlations between galaxies and their environment, we compare the direction of the projected angular momenta of CALIFA galaxies with the relative line-of-sight motions of their neighbor galaxies. To obtain sufficient statistical significance, we proceed by stacking individual kinematic maps along the directions of the angular momenta of CALIFA galaxies. Figure~\ref{stack} illustrates this procedure and displays the resulting composite map, where the direction of the angular momentum of CALIFA galaxies is systematically taken to be upward.
When defining `neighbor galaxies' of each CALIFA galaxy, we limit the difference of line-of-sight velocity to 500 km s$^{-1}$ and consider the projected distance up to 1 Mpc.

If the motions of neighbors are coherent with the rotation of CALIFA galaxies, the mean velocity of the galaxies in the right-side area will be positive (redshift) while that in the left-side area will be negative (blueshift) in the composite map. To check whether this trend exists, we first define the \emph{right-side} and \emph{left-side} in the composite map. We may simply divide the entire area into two sub-areas like Figure~\ref{cuts}(a) ({\it I-cut}), from which we can secure the most numerous neighbors.

In the I-cut, however, there are domains in which the influence of neighbors on the CALIFA galaxies is ambiguous. For example, suppose two neighbor galaxies that have the same line-of-sight velocities, masses and distances to the same CALIFA galaxy but are located at position angles of $1^{\circ}$ and $359^{\circ}$ from the upward (angular momentum vector) direction, respectively. They are actually very close to each other and thus their influence on the CALIFA galaxy may be almost the same.
However, in the I-cut, their influence will be oppositely counted, because they will belong to opposite sides.

The {\it X-cut}, as shown in Figure~\ref{cuts}(b), is an alternative selection designed to mitigate such a problem. In the X-cut, only the neighbor galaxies located between $45^{\circ}$ and $135^{\circ}$ or between $225^{\circ}$ and $315^{\circ}$ from the angular momentum vector direction are considered. Although the X-cut reduces the sample of neighbor galaxies into a half and thus may increase the statistical uncertainty, it is advantageous that the neighbors that are expected to hardly influence the current rotational directions of the CALIFA galaxies are excluded. Thus, hereafter, all of the results in this paper will be based on the X-cut.

When estimating the average line-of-sight motion of neighbor galaxies, we weight them by luminosity. In this weighting, again we have two options: absolute-luminosity (abs-L) weighting and relative-luminosity (rel-L) weighting.
In the abs-L weighting, the velocities of neighbors are weighted by their luminosities, regardless of the luminosity of their adjacent CALIFA galaxy. On the other hand, in the rel-L weighting, the velocities of neighbors are weighted by the luminosity \emph{ratio} between a neighbor galaxy and its adjacent CALIFA galaxy. The abs-L weighting is good for estimating the effect of \emph{absolute} environment regardless of the luminosity of CALIFA galaxies, while the rel-L weighting focuses on the direct interactions between the CALIFA galaxies and their neighbors. 

The selection between these two options depends on how the neighbor galaxies are related to the CALIFA galaxies.
That is, if the impact of a neighbor galaxy with a given mass is common to all CALIFA galaxies at the same distances regardless of their masses, the abs-L weighting must be used. On the other hand, if a neighbor more strongly influences less massive CALIFA galaxies, then the rel-L weighting may be appropriate.
Throughout this paper, we will present both of the results using the rel-L weighting and using the abs-L weighting. The comparison between those results will be useful to constrain the origin of dynamical coherence, if it exists.

\begin{figure*}[t]
\centering
\includegraphics[width=0.95\textwidth]{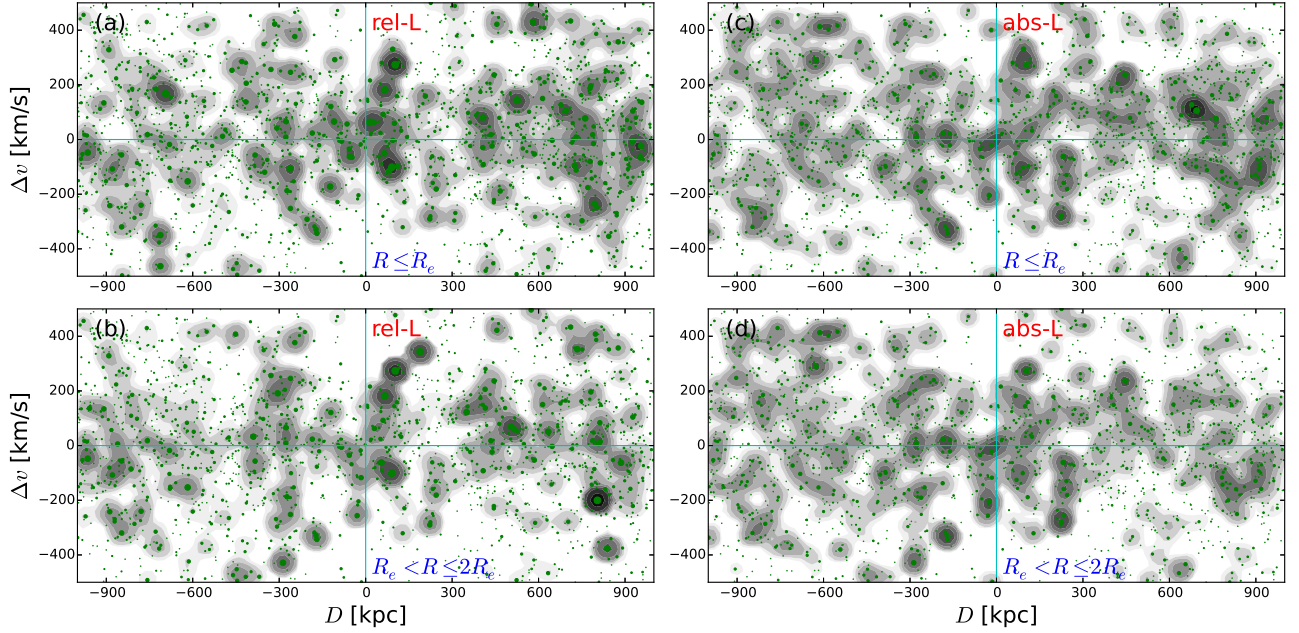}
\caption{(a) - (b) Relative-luminosity-weighted (rel-L-weighted) contour maps showing the neighbor galaxy distribution on the relative velocity versus projected distance plot, for the angular momenta measured (a) at $R{\le}R_e$ and (b) at $R_e<R{\le}2 R_e$ of individual CALIFA galaxies. The green dots denote the individual neighbor galaxies, and the size of each green dot represents the ratio of a neighbor luminosity to its adjacent CALIFA galaxy. Positive values in projected distance ($D$) are for the right-side neighbors, while negative values are for left-side neighbors in Figure~\ref{cuts}.
(c) - (d) The same as (a) - (b), but the contours are absolute-luminosity-weighted (abs-L-weighted) and the size of each green dot represents the neighbor luminosity itself, regardless of the luminosity of its adjacent CALIFA galaxy. In this figure and the figures hereafter, all results are based on the X-cut.\label{contour}}
\end{figure*}

\begin{figure*}[p]
\centering
\includegraphics[width=0.45\textwidth]{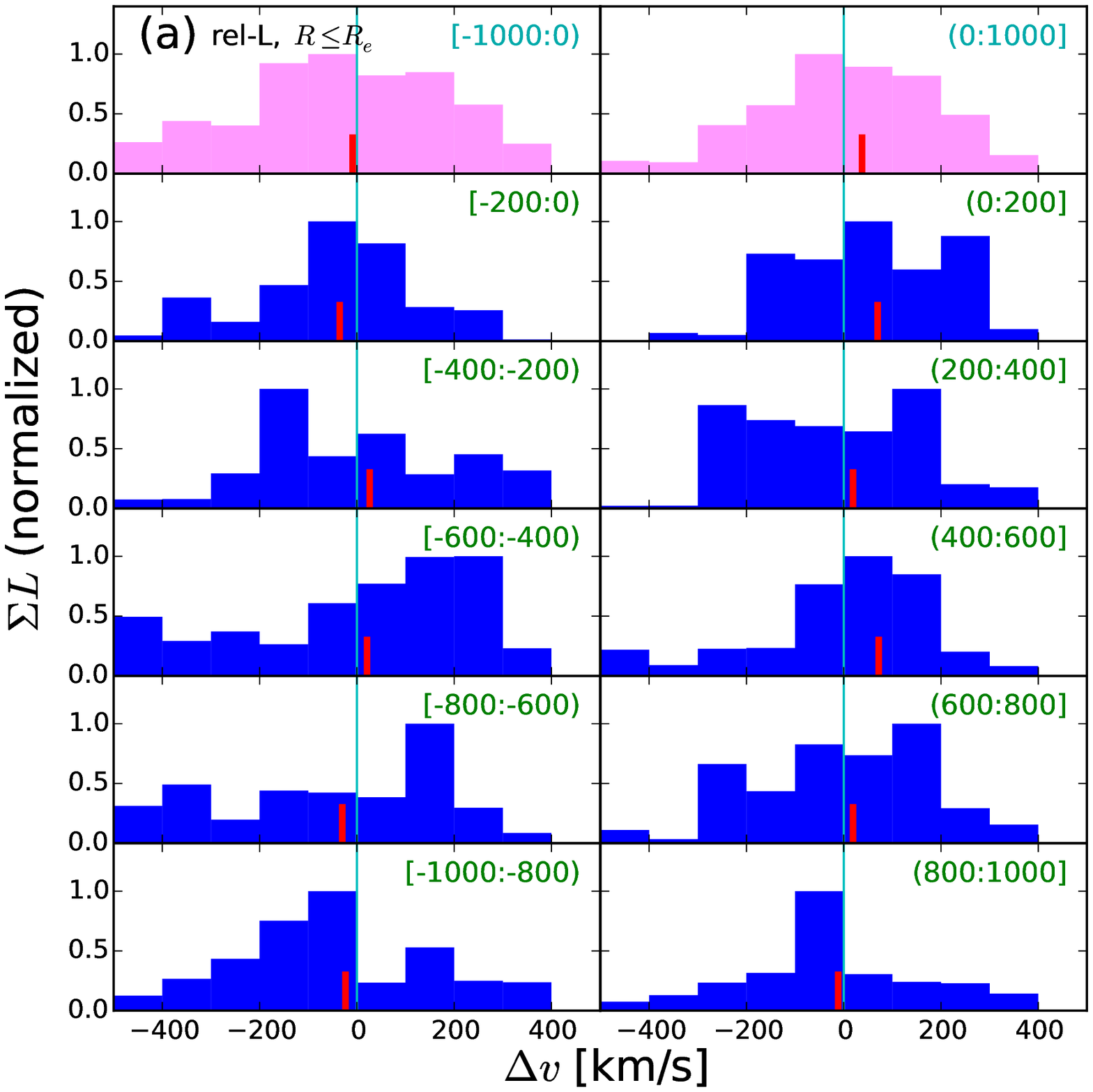}
\includegraphics[width=0.45\textwidth]{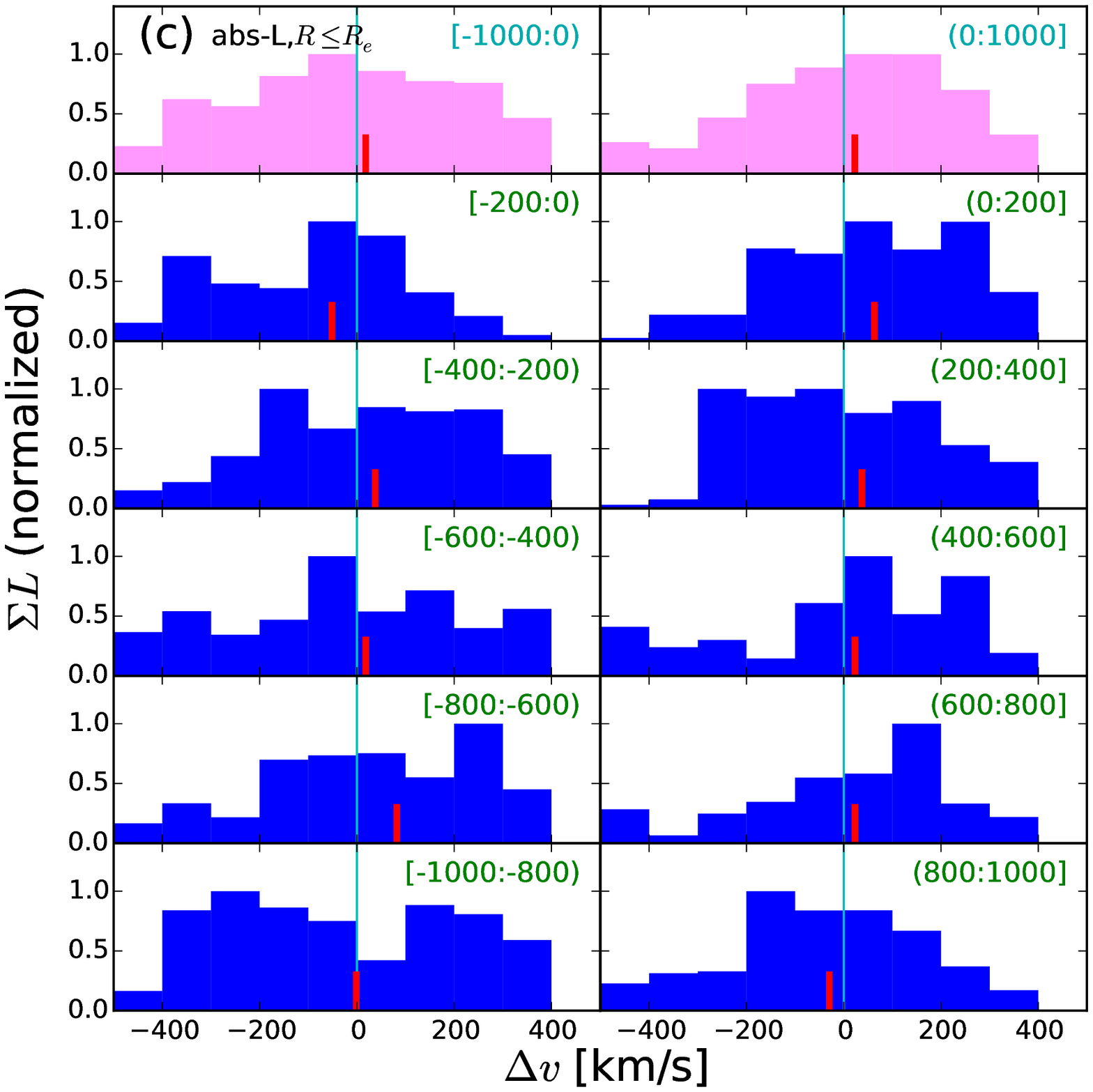}
\includegraphics[width=0.45\textwidth]{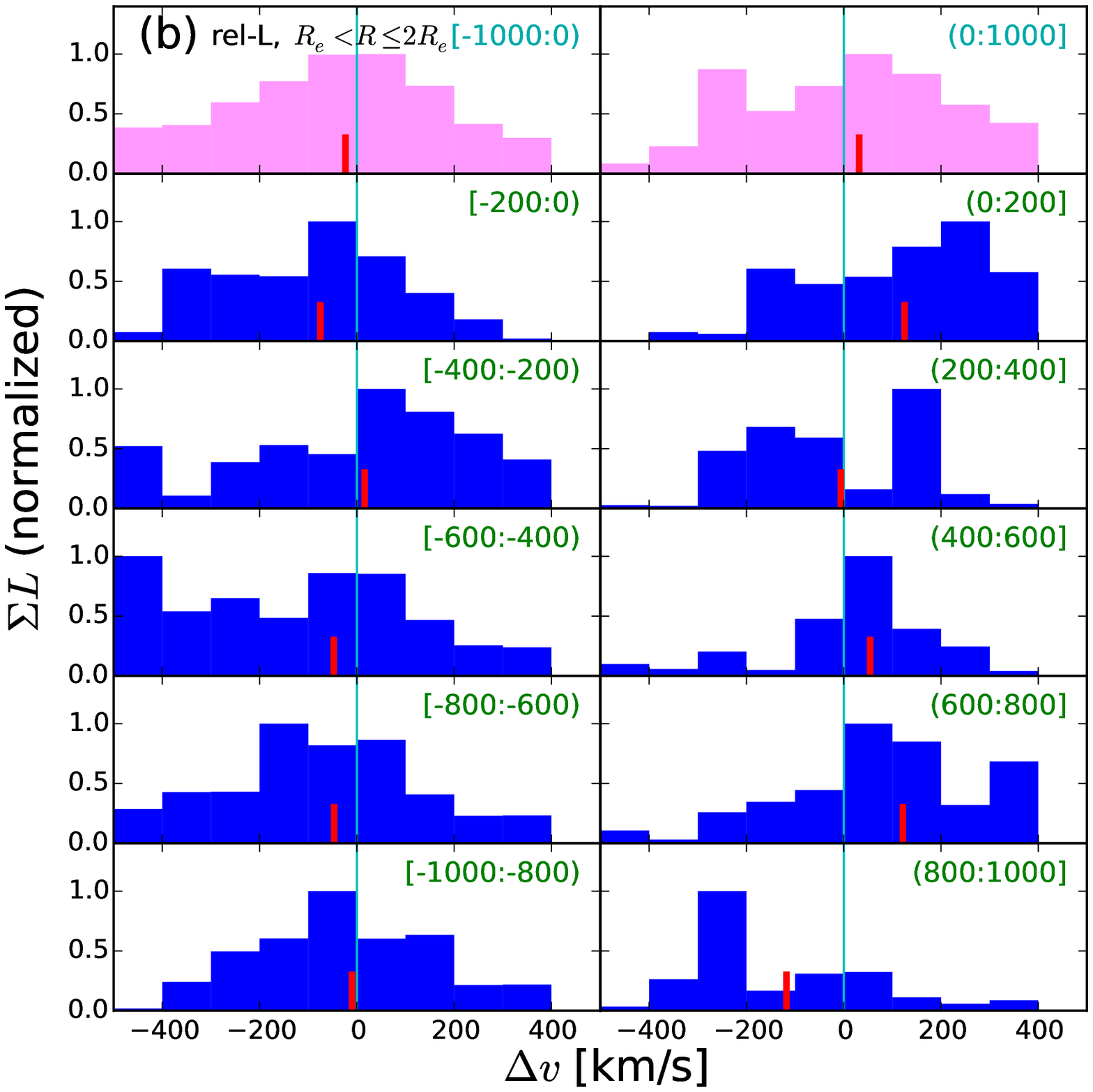}
\includegraphics[width=0.45\textwidth]{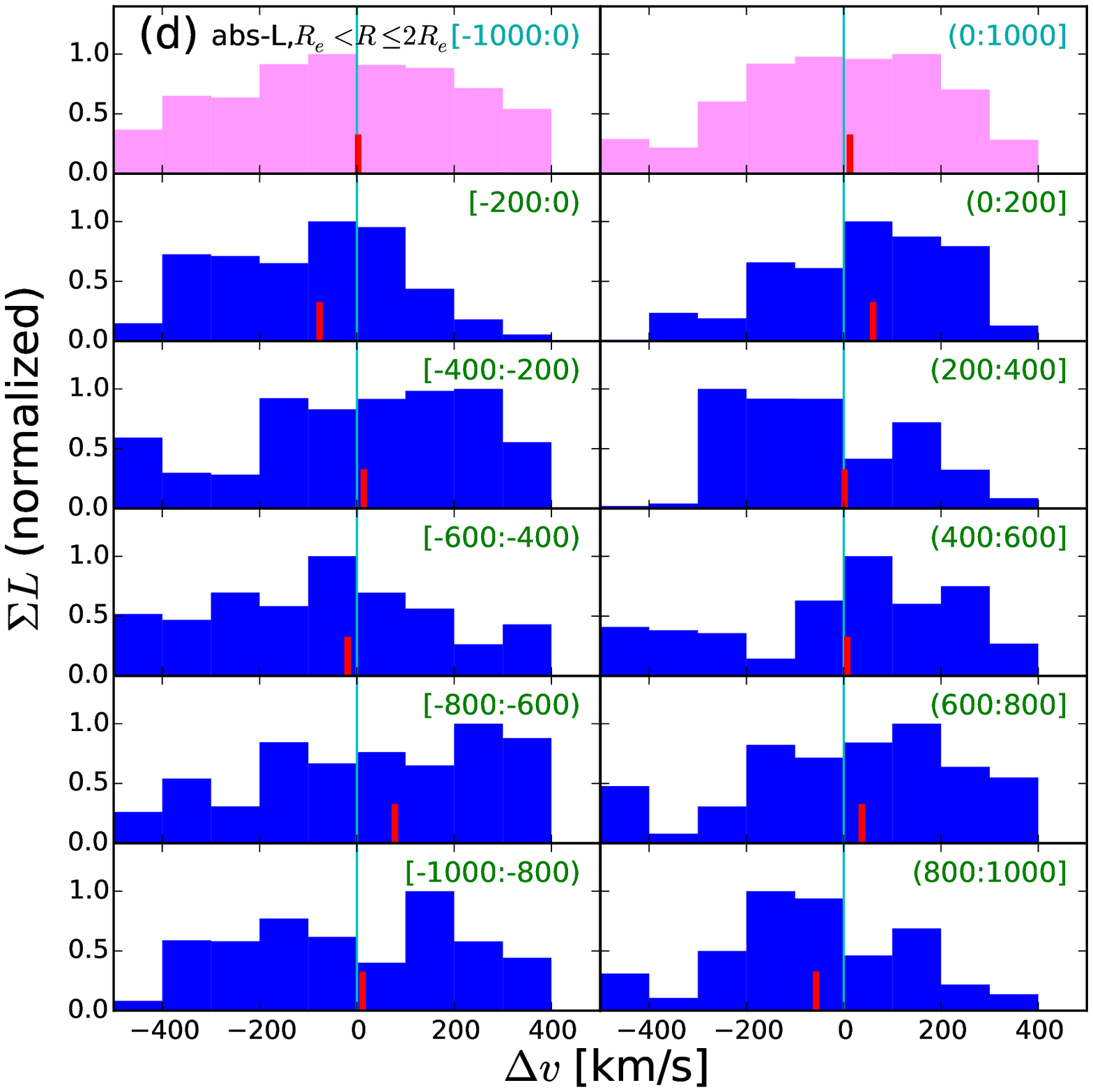}
\caption{The $\Delta v$ distributions at given projected distance ranges: (a) the rel-L-weighted neighbors for the angular momenta measured at $R{\le}R_e$, (b) the rel-L-weighted neighbors for the angular momenta measured at $R_e<R{\le}2R_e$, (c) the abs-L-weighted neighbors for the angular momenta measured at $R{\le}R_e$, and (d) the abs-L-weighted neighbors for the angular momenta measured at $R_e<R{\le}2R_e$. In each sub-panel, the histogram shows the normalized distribution of integrated luminosity of neighbors at given distance range that is denoted at the upper-right corner (in unit of kpc). The luminosity-weighted mean velocity is denoted by a red bar in each sub-panel.
\label{vrhist}}
\end{figure*}

\begin{figure*}[t]
\centering
\includegraphics[width=0.95\textwidth]{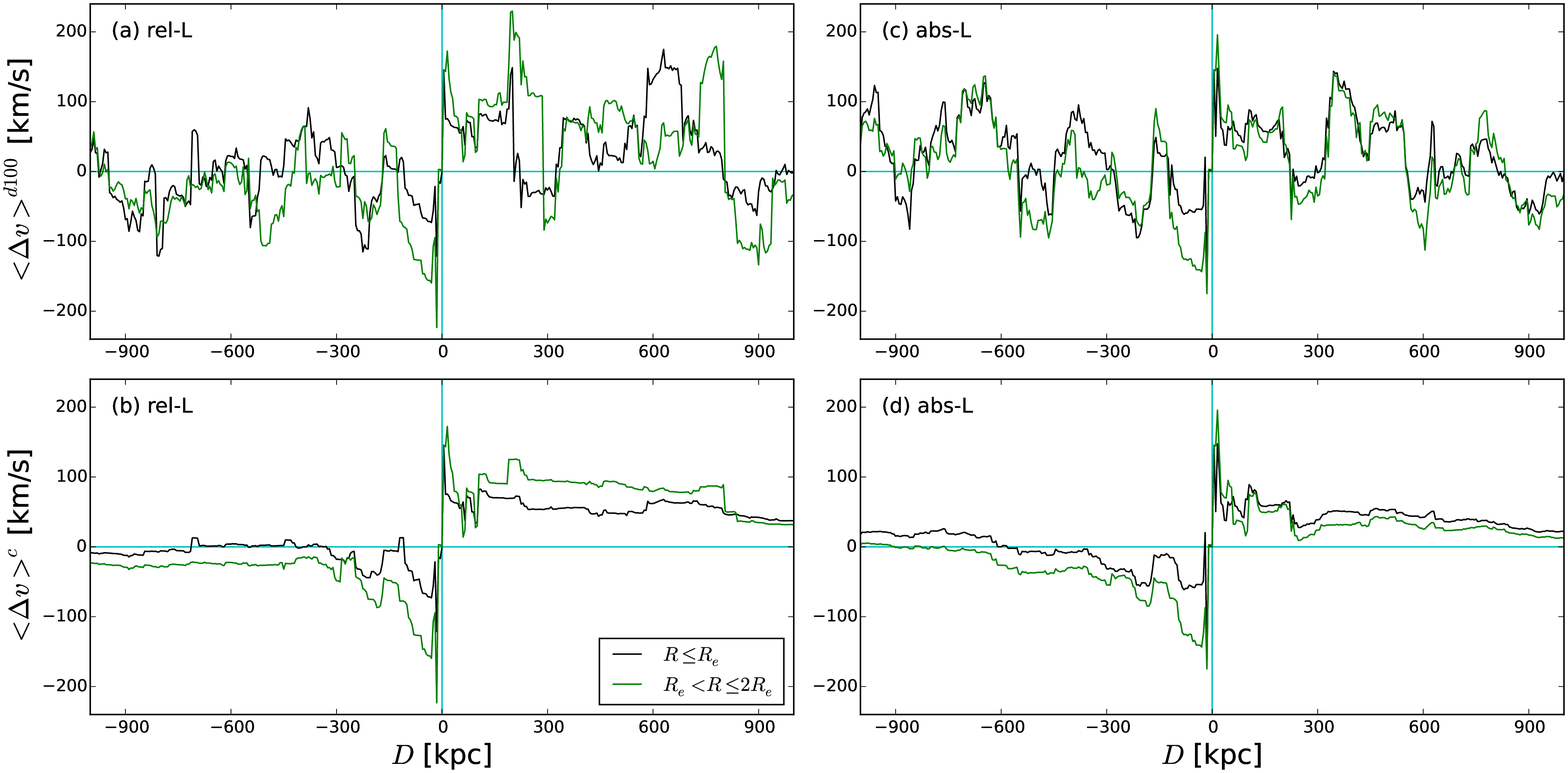}
\caption{(a) Radial profile of rel-L-weighted mean velocity. A smoothing kernel with 100-kpc size is applied. The black line shows the results based on angular momenta at $R{\le}R_e$, while the green line is based on angular momenta at $R_e<R\le 2R_e$. (b) Cumulative radial profile of rel-L-weighted mean velocity. (c) Radial profile of abs-L-weighted mean velocity. (d) Cumulative radial profile of abs-L-weighted mean velocity.\label{vprof}}
\end{figure*}

\section{RESULTS}\label{result}

\subsection{The Full Sample}\label{result1}

Before the quantitative investigation of the dynamical coherence, we present a qualitative analysis through the estimation of contour maps of the neighbor galaxy distribution on the relative line-of-sight velocity versus projected distance plot, as shown in Figure~\ref{contour}. These contour maps provide us an overall view on the dynamical distribution of neighbor galaxies around the CALIFA targets.
The contours show the log-scale number density distribution, weighted by the luminosity ratio of neighbor galaxy to CALIFA galaxy (rel-L-weighted; panels (a) and (b)) or weighted by the luminosity of neighbor galaxy (abs-L-weighted: panels (c) and (d)): darker contours represent higher-density regions. The number density for the contours was calculated at $100 \times 50$ grids and smoothed in a 3-grid scale with linear weighting.
Since some CALIFA galaxies have considerable amount of internal misalignment as shown in Figure~\ref{misalign}, the contour maps for the angular momenta at $R{\le}R_e$ and at $R_e<R\le 2R_e$ are not perfectly the same as each other, although some features are very similar (panels (a) versus (b), and (c) versus (d) in Figure~\ref{contour}).

Overall, we find a sequence of intense clumps from positive distance and positive relative velocity (right-side neighbors with redshift) to negative distance and negative relative velocity (left-side neighbors with blueshift), at a very small range of $|D|\lesssim200$ kpc, where $D$ is the projected distance to a given CALIFA galaxy. 
Such a sequence is apparently consistent with the coherence between the rotations of CALIFA galaxies and the motions of their neighbors, but some other clumps that do not agree with such coherence are also found. Furthermore, at $|D|>200$ kpc, it is not easy to visually recognize any signals of the dynamical coherence in the contours.

Note that the clumps in Figure~\ref{contour} may be largely affected by bright neighbors, because the contours are weighted by luminosity. Since more luminous/massive galaxies are expected to affect its neighbors more strongly, it is understandable that such clumpy features tend to show good coherence signals (particularly at small distances). However, this reminds us of the risk from small-number statistics: the entire trends may be dominated by a few luminous neighbors and coincidentally form apparent coherence signals. Thus, the coherence signals need to be further investigated in quantitative and statistical ways.

Figure~\ref{vrhist} shows the $\Delta v$ distribution of neighbor galaxies at given projected distance range. Although the neighbor galaxies are distributed over a wide range of $\Delta v$, the luminosity-weighted mean velocity tends to be slightly biased at some projected distance ranges. Such biases are particularly obvious at very close distances ($|D|{\leq}200$ kpc), in the context that the left-side neighbors tend to lean toward negative mean velocity, while the right-side neighbors tend to lean toward positive mean velocity. This can be regarded as the evidence of dynamical coherence between the rotational direction of CALIFA galaxies and the motions of neighbors. Such a trend is found even up to $|D|\approx800$ kpc for the rel-L-weighted results of outskirt ($R_e<R\leq 2R_e$) rotation. However, Figure~\ref{vrhist} does not show how statistically significant the mean velocity biases are.

To quantitatively assess the robustness of the apparent dynamical coherence, we define the luminosity-weighted mean velocity ($\langle\Delta v\rangle$) and plot the variation of $\langle\Delta v\rangle$ with projected distance to the CALIFA galaxies.
In Figure~\ref{vprof}, panels (a) and (c) show the derivative profiles, smoothed with a 100-kpc kernel, while panels (b) and (d) show the cumulative profiles.
In the derivative profiles, the mean velocity $\langle\Delta v\rangle^{d100}$ at given $D'$ is defined as:
\begin{equation}
 \langle\Delta v\rangle^{d100}(D') = \left\{ 
 \begin{array}{ll}
   \frac{\displaystyle \sum_{Rd(D',100)} \Delta v \mathcal{L}}{\displaystyle \sum_{Rd(D',100)} \mathcal{L}} & \textrm{if}\: D'>0 \\
   0 & \textrm{if}\: D'=0 \\
   \frac{\displaystyle \sum_{Ld(D',100)} \Delta v \mathcal{L}}{\displaystyle \sum_{Ld(D',100)} \mathcal{L}} & \textrm{if}\: D'<0 \textrm{,}
 \end{array} \right .
\end{equation}
where $\Delta v$ is the line-of-sight recession velocity of a neighbor galaxy relative to a given CALIFA galaxy, and the right-side distance range $Rd$ is:
\begin{equation}
 Rd(D',100) = \left\{ 
 \begin{array}{l}
   D'-100\,\textrm{kpc} < D \le D' \\
   \qquad\qquad\, \textrm{if}\: D'>100 \,\textrm{kpc}\\
   0 < D \le D' \\
   \qquad\qquad\, \textrm{if}\: 0<D'\le 100\,\textrm{kpc,}
 \end{array} \right .
\end{equation}
and the left-side distance range $Ld$ is:
\begin{equation}
 Ld(D',100) = \left\{ 
 \begin{array}{l}
   D' \le D < D'+100\,\textrm{kpc} \\
   \qquad\qquad \textrm{if}\: D'<-100 \,\textrm{kpc}\\
   D' \le D < 0 \\
   \qquad\qquad \textrm{if}\: -100\,\textrm{kpc}\le D'<0 \textrm{.}
 \end{array} \right .
\end{equation}
On the other hand, in the cumulative profiles, the mean velocity $\langle\Delta v\rangle^{c}$ at given $D'$ is defined as:
\begin{equation}
 \langle\Delta v\rangle^{c}(D') = \left\{ 
 \begin{array}{ll}
   \frac{\displaystyle \sum_{0<D\le D'} \Delta v \mathcal{L}}{\displaystyle \sum_{0<D\le D'} \mathcal{L}} & \textrm{if}\: D'>0 \\
   0 & \textrm{if}\: D'=0 \\
   \frac{\displaystyle \sum_{D'\le D<0} \Delta v \mathcal{L}}{\displaystyle \sum_{D'\le D<0} \mathcal{L}} & \textrm{if}\: D'<0 \textrm{.}
 \end{array} \right .
\end{equation}
$\mathcal{L}$ is the luminosity ratio of a neighbor galaxy to a given CALIFA galaxy (= $L_{\textrm{\tiny neighbor}}/L_{\textrm{\tiny CALIFA}}$) when the rel-L weighting is applied, while it is simply the luminosity of a neighbor galaxy ($L_{\textrm{\tiny neighbor}}$) for the abs-L weighting.
Note that the derivative profiles ($\langle\Delta v\rangle^{d100}$) and the cumulative profiles ($\langle\Delta v\rangle^c$) coincide with each other at $|D|{\le}100$ kpc.

\begin{figure*}[t]
\centering
\includegraphics[width=0.95\textwidth]{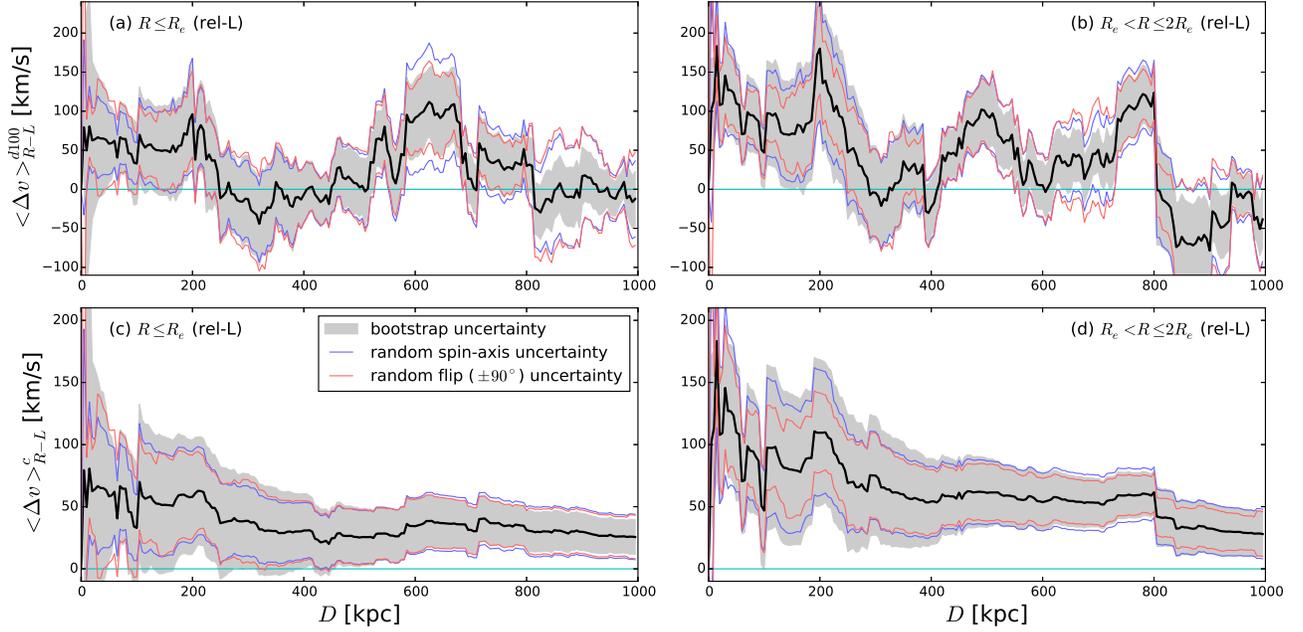}
\caption{Right-left-merged radial profiles (thick black lines) of rel-L-weighted mean velocity for angular momenta measured (a) at $R{\le}R_e$ and (b) at $R_e<R\le 2R_e$. (c) Cumulative version of (a). (d) Cumulative version of (b). The bootstrap uncertainty (shaded areas), random spin-axis uncertainty (blue lines) and random flip ($\pm90^{\circ}$) uncertainty (red lines) are also denoted (see the main text for the detailed description of the uncertainties).\label{vprofbs}}
\end{figure*}

\begin{figure*}[t]
\centering
\includegraphics[width=0.95\textwidth]{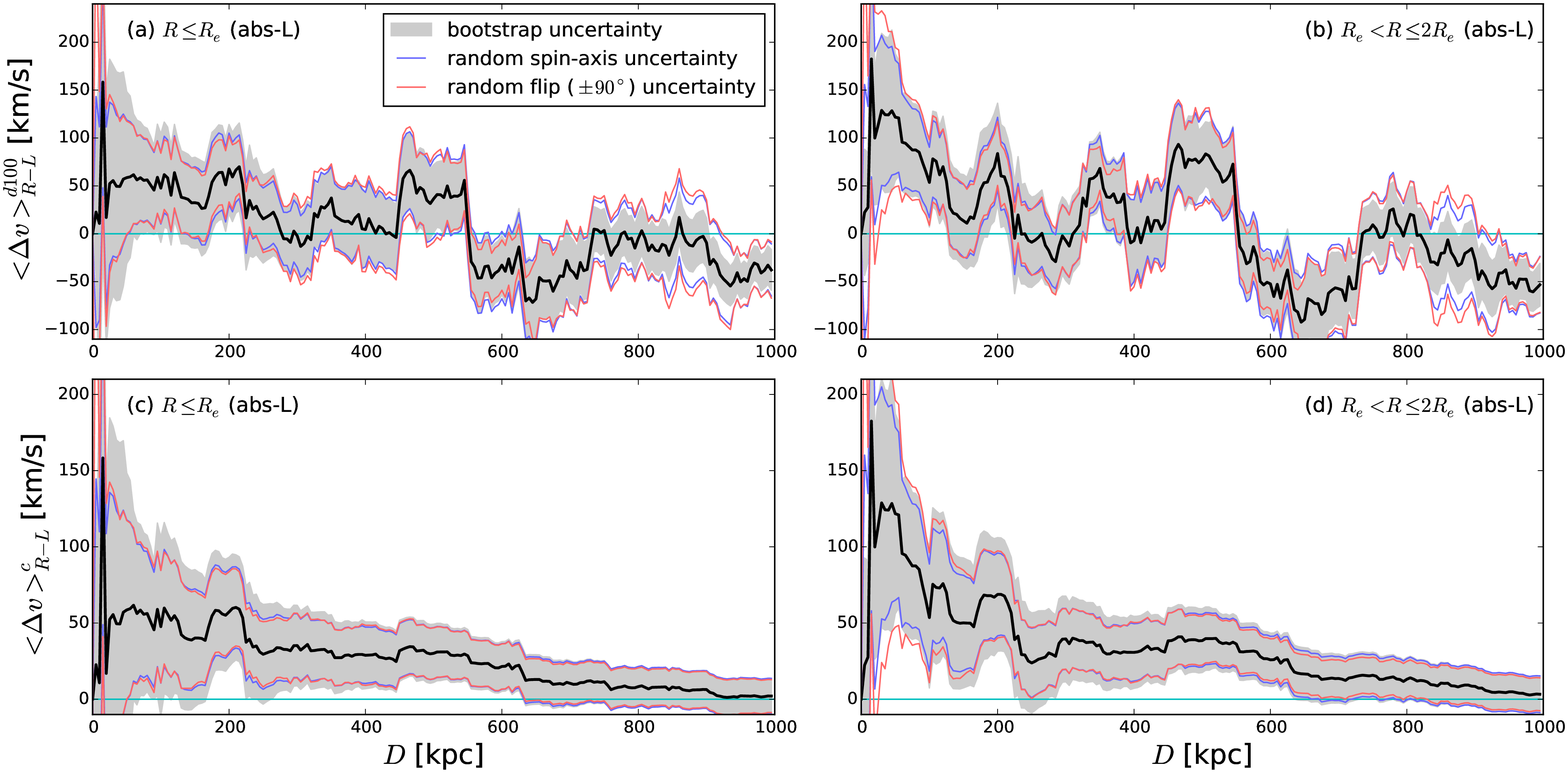}
\caption{The same as Figure~\ref{vprofbs}, except that the abs-L weighting is applied. \label{vprofbs2}}
\end{figure*}

In Figure~\ref{vprof}(a) and (c), the derivative profiles are a little bit noisy, but some coherence signals are visible commonly for the rel-L weighting and the abs-L weighting: the mean velocity tends to be positive at the right-side and negative at left-side, which is particularly conspicuous at $|D|\lesssim 300$ kpc. In Figure~\ref{vprof}(b) and (d), the cumulative profiles are smoother than the derivative ones, showing the coherence signals more obviously.

One notable feature in Figure~\ref{vprof}(b) and (d) is its asymmetry: the mean velocity of right-side neighbors converges to zero more lately than that of left-side neighbors does. For this apparent asymmetry, as well as coincidence due to small-number statistics, a selection bias may be partially responsible. The CALIFA survey targeted apparently bright galaxies (Figure~\ref{spdist}(b)). If galaxies in the Universe were homogeneously distributed along redshift, such target selection might not cause a bias. However, in reality, many galaxies tend to gather around large-scale structures: clusters, groups, filaments and walls. If we observe towards such gathering of galaxies, then the selection of bright targets (in apparent magnitude) may be biased to galaxies at lower redshifts, which means that the mean redshifts of their neighbors (not selected as targets) may tend to be higher (= more receding). If the relative line-of-sight velocities of the neighbors selected within $\pm500$ km s$^{-1}$ are dominated by their peculiar motions rather than the Hubble flow (e.g., in a galaxy cluster), this effect may be negligible, but our targets are mostly in non-cluster environments. That is, the asymmetry in Figure~\ref{vprof} may be, at least partially, because apparently brighter galaxies (preferentially at lower redshifts) had higher chances to be selected as the CALIFA targets.

\begin{deluxetable*}{ccrrrr}
\tablenum{1} \tablecolumns{6} \tablecaption{Coherence Signals (All CALIFA Galaxies)} \tablewidth{0pt}
\tablehead{ Weighting & Angular & $D$ & $\langle\Delta v\rangle^{c}_{R-L} \pm\sigma_{\textrm{\tiny BST}}$ & $\pm\sigma_{\textrm{\tiny RAX}}$ & $\pm\sigma_{\textrm{\tiny RFA}}$ \\
& momentum & [kpc] & [km s$^{-1}$] & [km s$^{-1}$] & [km s$^{-1}$] }
\startdata
Rel-L & $R\leq R_e$ & 30 & $66.6\pm87.2\,(0.8\sigma)$ & $\pm55.3\,(1.2\sigma)$ & $\pm74.2\,(0.9\sigma)$ \\
 & & 200 & $57.9\pm46.0\,(1.3\sigma)$ & $\pm36.8\,(1.6\sigma)$ & $\pm33.6\,(1.7\sigma)$ \\
 & & 500 & $25.2\pm23.4\,(1.1\sigma)$ & $\pm21.8\,(1.2\sigma)$ & $\pm22.1\,(1.1\sigma)$ \\
 & & 800 & $31.1\pm15.6\,(2.0\sigma)$ & $\pm18.7\,(1.7\sigma)$ & $\pm15.8\,(2.0\sigma)$ \\
\hline
Rel-L & $R_e<R\leq 2R_e$ & 30 & $145.7\pm55.5\,(2.6\sigma)$ & $\pm70.4\,(2.1\sigma)$ & $\pm49.9\,(2.9\sigma)$ \\
 & & 200 & $109.5\pm58.9\,(1.9\sigma)$ & $\pm50.6\,(2.2\sigma)$ & $\pm32.1\,(3.4\sigma)$ \\
 & & 500 & $61.1\pm27.8\,(2.2\sigma)$ & $\pm25.6\,(2.4\sigma)$ & $\pm22.8\,(2.7\sigma)$ \\
 & & 800 & $61.7\pm17.6\,(3.5\sigma)$ & $\pm20.2\,(3.0\sigma)$ & $\pm15.5\,(4.0\sigma)$ \\
\hline
Abs-L & $R\leq R_e$ & 30 & $55.6\pm123.6\,(0.4\sigma)$ & $\pm77.7\,(0.7\sigma)$ & $\pm81.9\,(0.7\sigma)$ \\
 & & 200 & $57.3\pm35.9\,(1.6\sigma)$ & $\pm27.7\,(2.1\sigma)$ & $\pm26.4\,(2.2\sigma)$ \\
 & & 500 & $30.1\pm19.1\,(1.6\sigma)$ & $\pm16.8\,(1.8\sigma)$ & $\pm16.7\,(1.8\sigma)$ \\
 & & 800 & $8.3\pm12.8\,(0.6\sigma)$ & $\pm13.2\,(0.6\sigma)$ & $\pm12.3\,(0.7\sigma)$ \\
\hline
Abs-L & $R_e<R\leq 2R_e$ & 30 & $128.9\pm81.8\,(1.6\sigma)$ & $\pm76.1\,(1.7\sigma)$ & $\pm104.8\,(1.2\sigma)$ \\
 & & 200 & $68.8\pm37.3\,(1.8\sigma)$ & $\pm27.2\,(2.5\sigma)$ & $\pm28.6\,(2.4\sigma)$ \\
 & & 500 & $38.5\pm20.9\,(1.8\sigma)$ & $\pm17.3\,(2.2\sigma)$ & $\pm16.1\,(2.4\sigma)$ \\
 & & 800 & $13.1\pm14.2\,(0.9\sigma)$ & $\pm13.5\,(1.0\sigma)$ & $\pm11.7\,(1.1\sigma)$
\enddata
\tablecomments{BST = Bootstrap uncertainty. RAX = Random spin-axis uncertainty. RFA = Randomly flipped ($\pm90^{\circ}$) spin-axis uncertainty.}
\label{signals}
\end{deluxetable*}

In the mean velocity profiles, both of the positive values for the right-side neighbors ($D>0$) and the negative values for the left-side neighbors ($D<0$) support the coherence between galaxy rotation and neighbor motions.
Thus, we can more simplify Figure~\ref{vprof} by defining the \emph{right-left-merged} mean velocities, as follows:
\begin{equation}
 \langle\Delta v\rangle^{d100}_{R-L}(D') = \frac{\displaystyle \Bigg(\sum_{Rd(D',100)} \Delta v \mathcal{L} \Bigg) - \Bigg(\sum_{Ld(-D',100)} \Delta v \mathcal{L} \Bigg)}{\displaystyle \Bigg( \sum_{Rd(D',100)} \mathcal{L} \Bigg) + \Bigg(\sum_{Ld(-D',100)} \mathcal{L} \Bigg)}
\end{equation}
and
\begin{equation}
 \langle\Delta v\rangle^{c}_{R-L}(D') = \frac{\displaystyle \Bigg(\sum_{0<D{\le}D'} \Delta v \mathcal{L} \Bigg) - \Bigg(\sum_{-D'{\le}D<0} \Delta v \mathcal{L} \Bigg)}{\displaystyle \Bigg( \sum_{0<D{\le}D'} \mathcal{L} \Bigg) + \Bigg(\sum_{-D'{\le}D<0} \mathcal{L} \Bigg)}
\end{equation}
where $D'>0$.
Roughly described, the right-left-merged mean velocity corresponds to the right-side mean velocity ($D>0$) subtracted by left-side mean velocity ($D<0$) in Figure~\ref{vprof}. Thus, the right-left-merged mean velocity must be positive, if there is the dynamical coherence between galaxy rotation and neighbor motions. In these right-left-merged profiles, the CALIFA target selection bias mentioned in the previous paragraph is canceled out on average.

\begin{figure*}[p]
\centering
\includegraphics[width=0.95\textwidth]{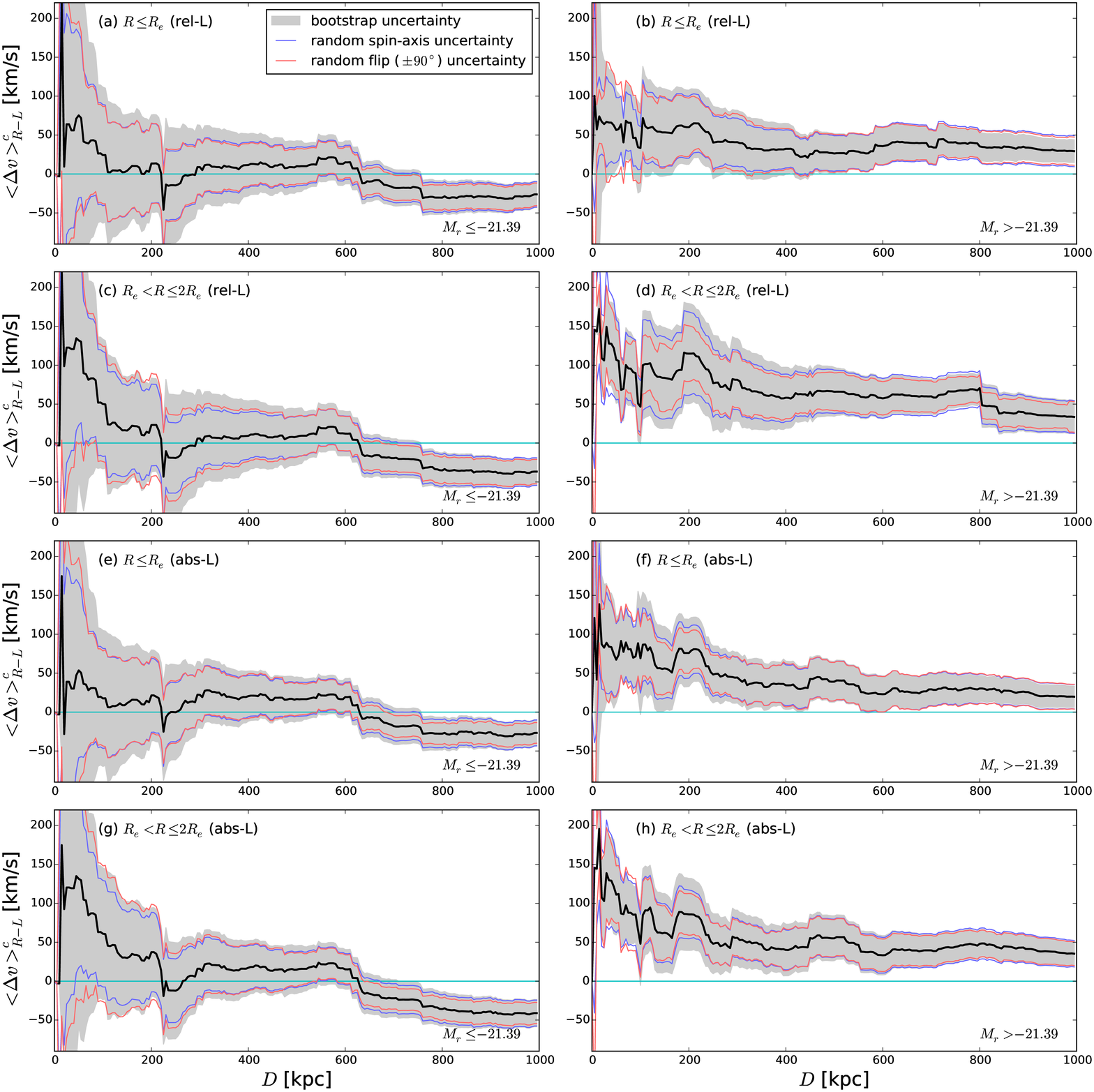}
\caption{Right-left-merged and cumulative profiles of luminosity-weighted mean velocity, (a)(c)(e)(g) for the subsample with high luminosity ($M_r{\le}-21.39$), and (b)(d)(f)(h) for the subsample with low luminosity ($M_r>-21.39$). (a) - (d) show the rel-L-weighted results, while (e) - (h) show the abs-L-weighted results.\label{vprofmagbs}}
\end{figure*}

\begin{figure*}[p]
\centering
\includegraphics[width=0.95\textwidth]{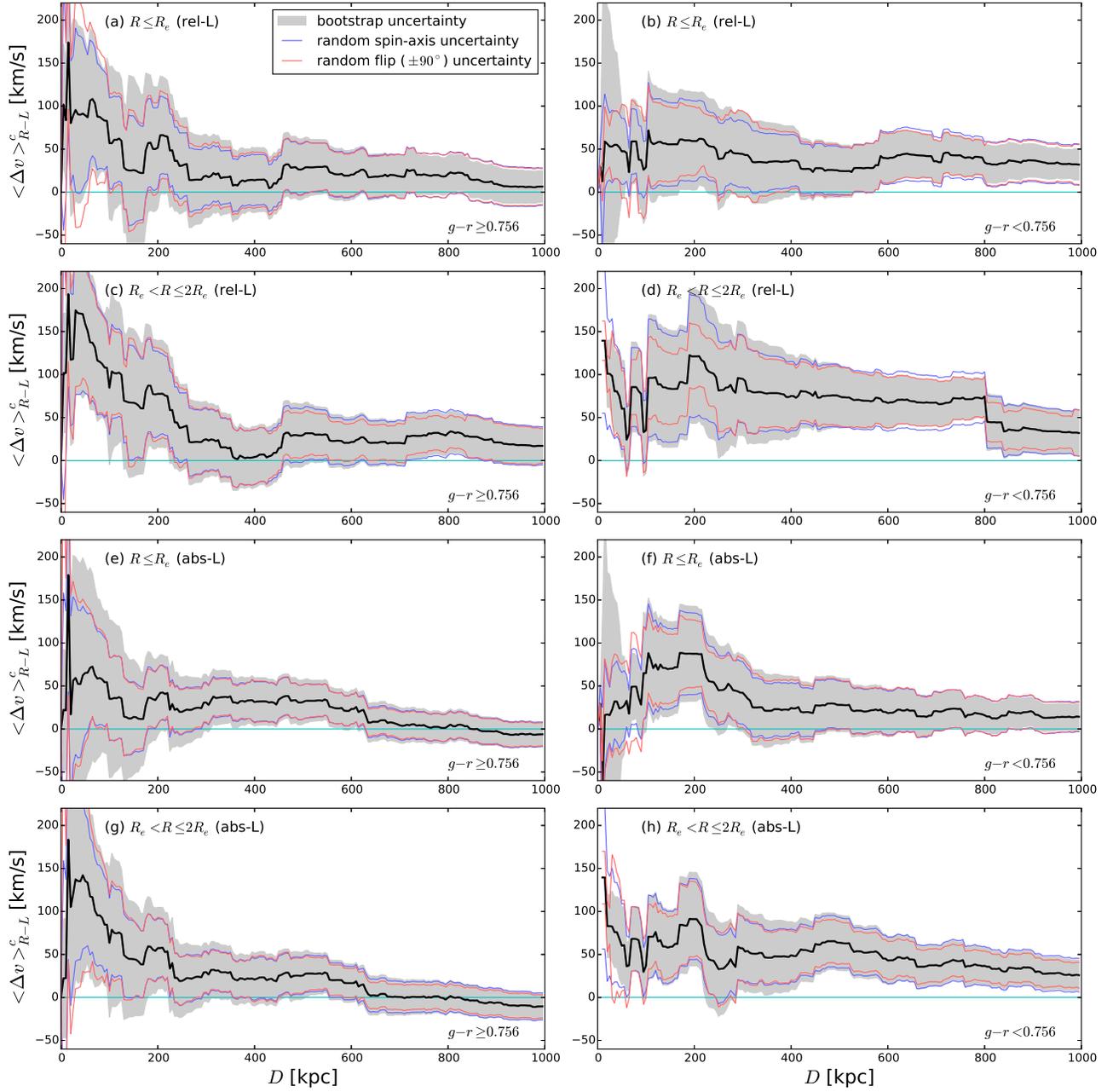}
\caption{Right-left-merged and cumulative profiles of luminosity-weighted mean velocity, (a)(c)(e)(g) for the subsample with red color ($g-r{\ge}0.756$), and (b)(d)(f)(h) for the subsample with blue color ($g-r<0.756$). \label{vprofcolbs}}
\end{figure*}

\begin{figure*}[p]
\centering
\includegraphics[width=0.95\textwidth]{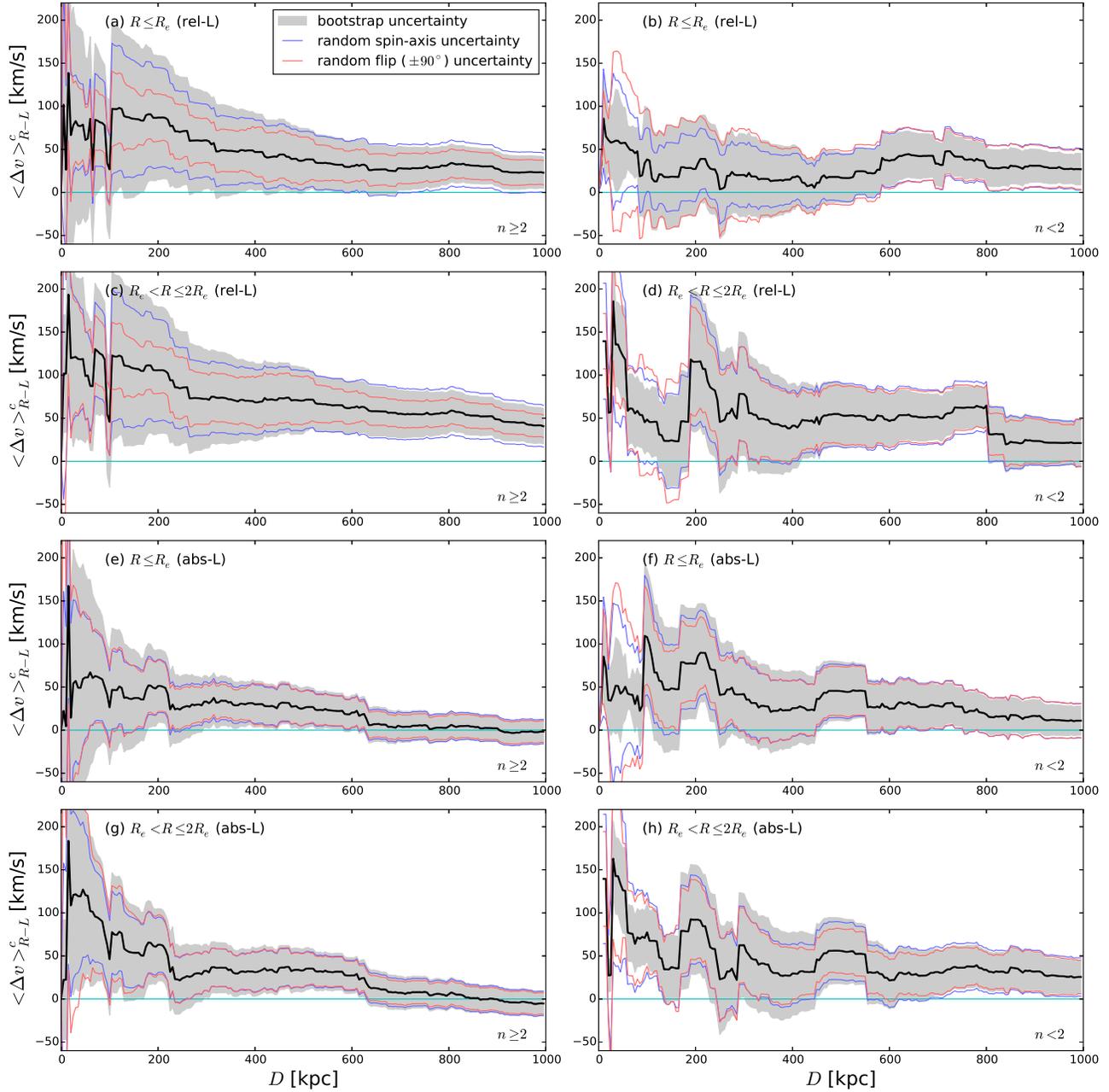}
\caption{Right-left-merged and cumulative profiles of luminosity-weighted mean velocity, (a)(c)(e)(g) for the subsample with large S{\'e}rsic index ($n{\ge}2$), and (b)(d)(f)(h) for the subsample with small S{\'e}rsic index ($n<2$). \label{vprofserbs}}
\end{figure*}

\begin{figure*}[p]
\centering
\includegraphics[width=0.95\textwidth]{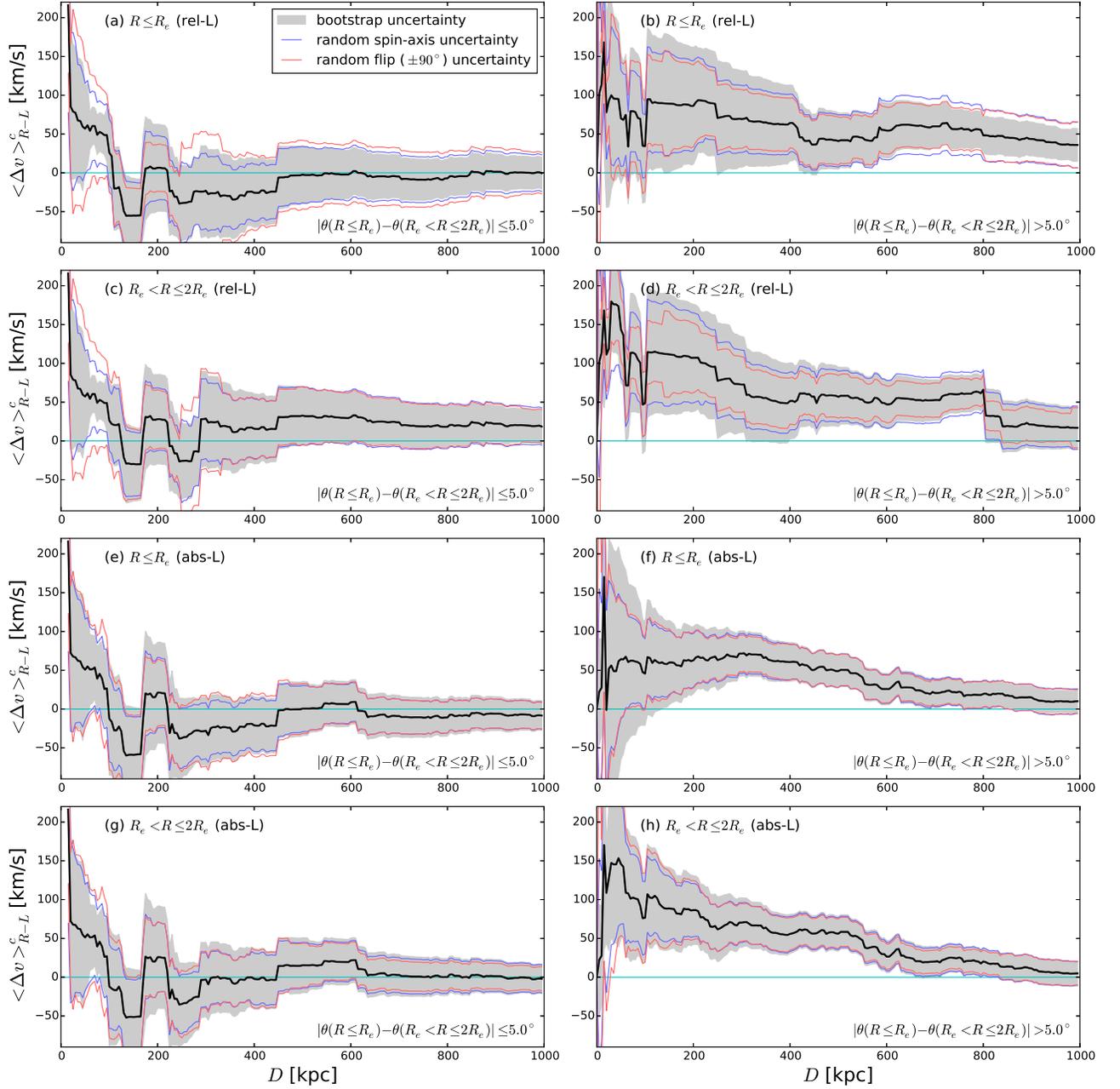}
\caption{Right-left-merged and cumulative profiles of luminosity-weighted mean velocity, (a)(c)(e)(g) for the subsample with small internal misalignment ($|\theta(R{\le}R_e)-\theta(R_e<R\le 2R_e)|{\le}5.0^{\circ}$), and (b)(d)(f)(h) for the subsample with large internal misalignment ($|\theta(R{\le}R_e)-\theta(R_e<R\le 2R_e)|>5.0^{\circ}$). \label{vprofmisbs}}
\end{figure*}

\begin{figure*}[p]
\centering
\includegraphics[width=0.95\textwidth]{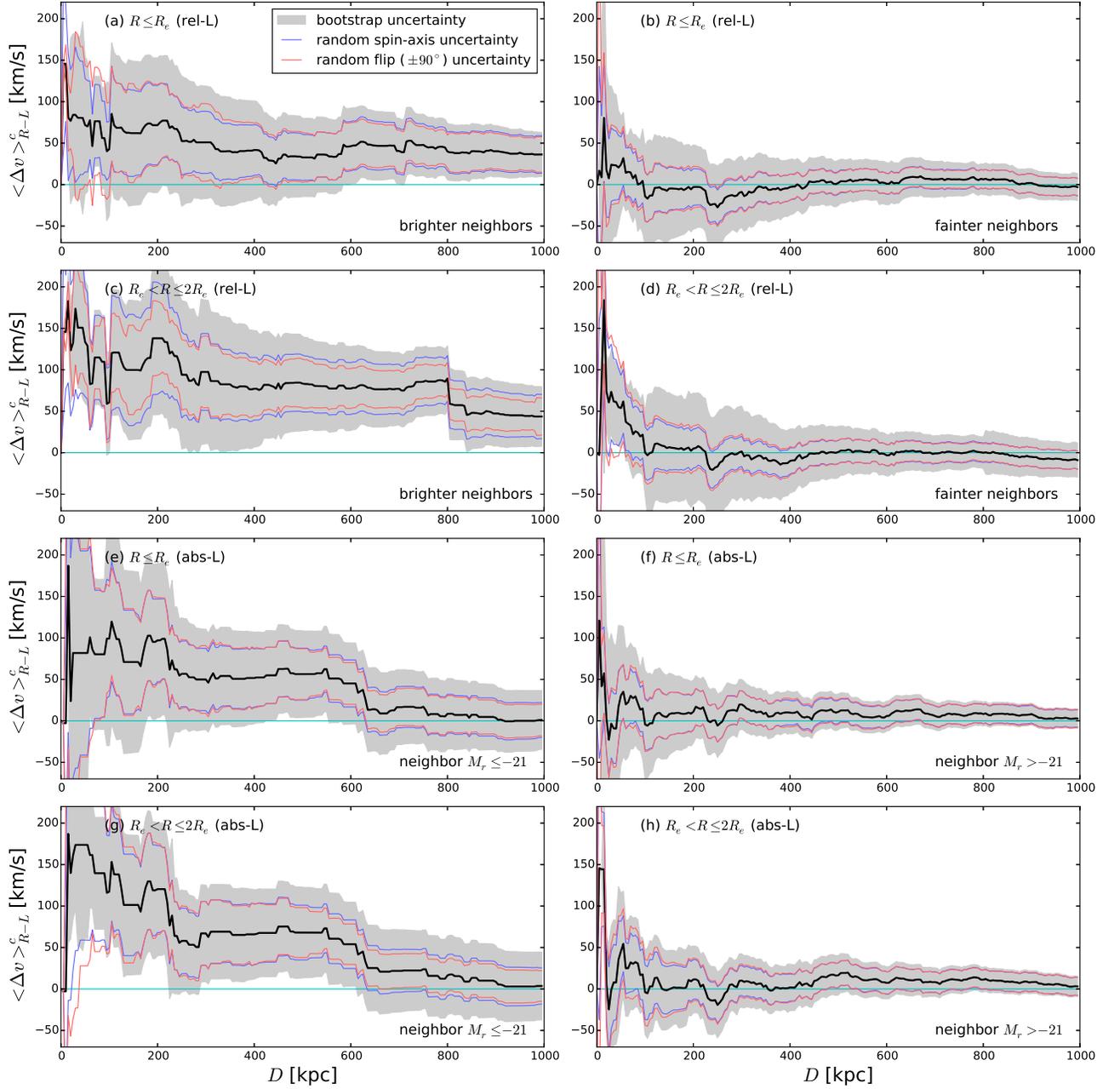}
\caption{With the full sample of CALIFA galaxies, right-left-merged and cumulative profiles of luminosity-weighted mean velocity, (a)(c)(e)(g) for the bright neighbors (brighter than the adjacent CALIFA galaxy or brighter than $M_r=-21$), and (b)(d)(f)(h) for the faint neighbors (fainter than the adjacent CALIFA galaxy or $M_r=-21$). \label{neimag}}
\end{figure*}

\begin{figure*}[p]
\centering
\includegraphics[width=0.95\textwidth]{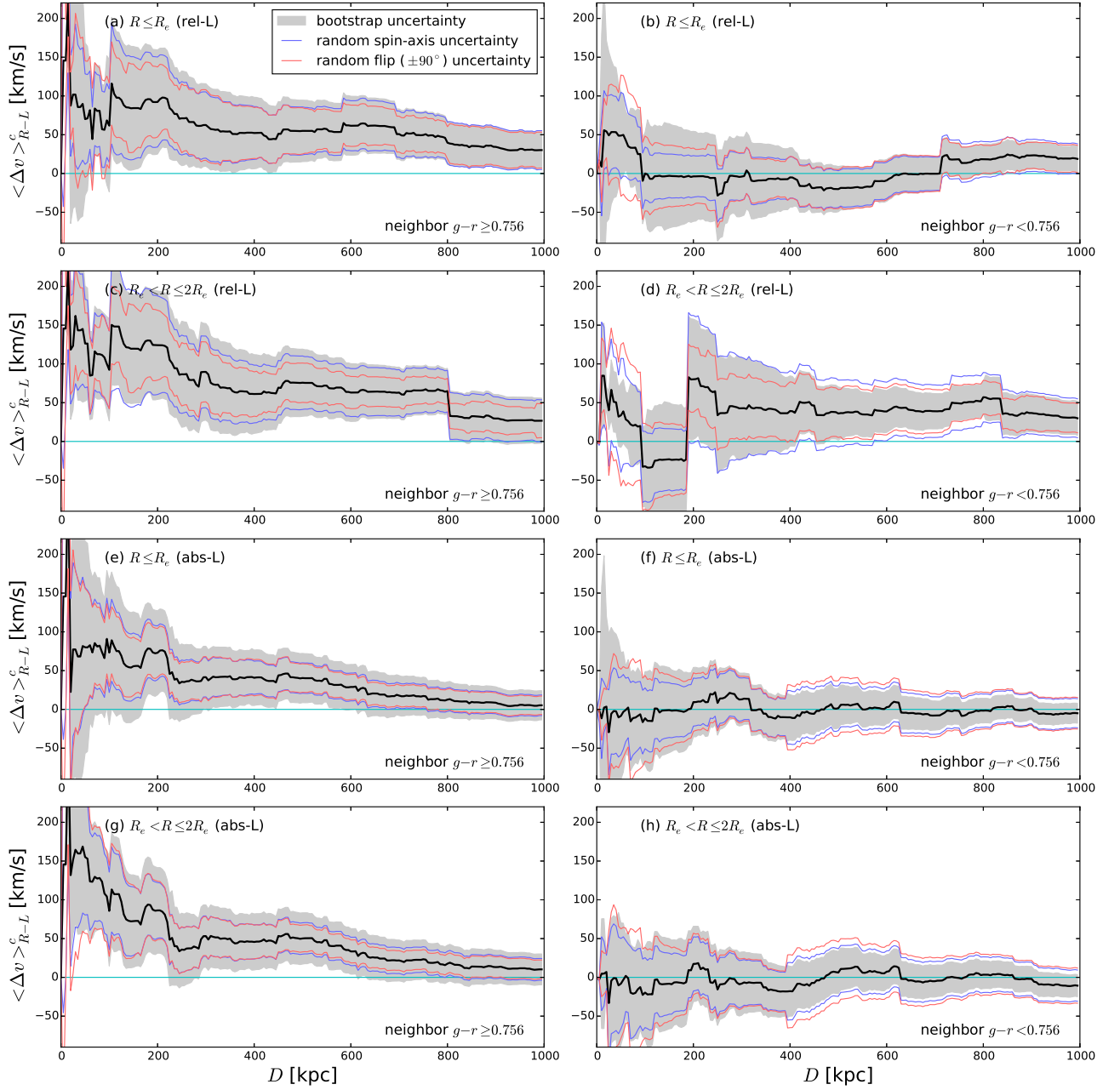}
\caption{With the full sample of CALIFA galaxies, right-left-merged and cumulative profiles of luminosity-weighted mean velocity, (a)(c)(e)(g) for the red neighbors ($g-r\geq 0.756$), and (b)(d)(f)(h) for the faint neighbors ($g-r<0.756$).  \label{neicol}}
\end{figure*}

Figures~\ref{vprofbs} and \ref{vprofbs2} show the right-left-merged mean velocity profiles. These merged profiles more efficiently present how significant the coherence signals are. For quantitative estimation of the significance of the coherence signals, we try three different statistical methods.
The first is the bootstrap (BST) uncertainty: the neighbor galaxies were randomly resampled with replacement, and the standard deviation of the results from 1000-times resampling experiments was estimated. The BST uncertainty is denoted by shaded areas in Figures~\ref{vprofbs} - \ref{neicol}.
The BST uncertainty works well in most cases, but caution is necessary when the number of neighbors is too small (e.g., at very close distances from the CALIFA galaxies). For example, the number of neighbors at $D\leq 30$ kpc is 27 and 24 for central and outskirt rotations, respectively, which is not too small to estimate BST uncertainty. However, this number can be substantially reduced when we use some subsamples. Other kinds of uncertainty may be helpful in those cases.

The second is the random spin-axis (RAX) test for the CALIFA galaxies.
The null hypothesis for this test is that ``the spin axis of each CALIFA galaxy is randomly determined regardless of the motions of its neighbor galaxies''. In this test, we can examine how easily the apparent `dynamical coherence' can be reproduced by random assignment of rotations to the CALIFA galaxies, for the given distribution of neighbor motions.
The RAX uncertainty is calculated by the following procedure:
\begin{enumerate}
 \item[1.] Assign a random vector to each CALIFA galaxy (replacing its original angular momentum vector).
 \item[2.] Align the neighbor galaxies around each CALIFA galaxy for the randomly assigned vector direction to be upward.
 \item[3.] Build a composite map of the newly aligned neighbor galaxies like Figure~\ref{stack}(e).
 \item[4.] Estimate the $\langle\Delta v\rangle^{d100}_{R-L}$ or $\langle\Delta v\rangle^{c}_{R-L}$ profiles using the new composite map.
 \item[5.] Repeat the steps 1 -- 4 by 1000 times and calculate the root-mean-squared deviation of the $\langle\Delta v\rangle^{d100}_{R-L}$ or $\langle\Delta v\rangle^{c}_{R-L}$ profiles.
\end{enumerate}
The RAX uncertainty is denoted by blue lines in Figures~\ref{vprofbs} - \ref{neicol}.

The last is the randomly flipped spin-axis (RFA) uncertainty. This test is similar to the RAX test, but the spin axis of each CALIFA galaxy is randomly flipped by $+90^{\circ}$ or $-90^{\circ}$ instead of being perfectly randomized. If it is true that there is some coherence between the CALIFA galaxy rotation and its neighbor motions, the RFA test will return smaller uncertainty (= higher statistical significance), because the neighbors in the X-cut regions after the random flipping by $\pm90^{\circ}$ have genuinely random motions, not contaminated by coherent motions. The RFA uncertainty is denoted by red lines in Figures~\ref{vprofbs} - \ref{neicol}. In the results, RFA uncertainty actually tends to be smaller than RAX uncertainty when the coherence signals are considerable (e.g., $>2\sigma_{\textrm{\tiny RAX}}$; but not always), which indirectly supports the existence of dynamical coherence.

In Figures~\ref{vprofbs} and \ref{vprofbs2}, there are some noticeable features. First, both in the derivative and cumulative profiles, significant coherence signals are detected for the outskirt angular momenta ($R_e<R\le 2R_e$; up to $3.5\sigma_{\textrm{\tiny BST}}$, $3.0\sigma_{\textrm{\tiny RAX}}$ and $4.0\sigma_{\textrm{\tiny RFA}}$), whereas the signals are hardly significant for the central angular momenta ($R\le R_e$). Such a trend is obvious when the rel-L weighting is applied (Figure~\ref{vprofbs}). However, in the case of the abs-L weighting (Figure~\ref{vprofbs2}), the coherence signals are marginal to any uncertainty. Second, for the outskirt angular momenta ($R_e<R\le 2R_e$), a few strong peaks are noted in the derivative profile. The individual $D$ values for those peaks may not be universal and may change if we use different samples. However, the wide range of the peak loci shows that the neighbors at various distances up to 800~kpc have dynamical coherence with the CALIFA galaxies.
Finally, at $D>800$ kpc, the coherence signals are insignificant.
The $\langle\Delta v\rangle^{c}_{R-L}$ values, their statistical uncertainties and the corresponding confidence levels at 30, 200, 500, and 800 kpc distances are listed in Table~\ref{signals}.

\subsection{Various Subsamples}\label{result2}

In order to understand more detailed aspects of the coherence between galaxy rotation and neighbor motions, we additionally conducted the same work as Section~\ref{result1} for several sets of subsamples divided by various criteria: luminosity, color, S{\'e}rsic index and internal kinematic misalignment of the CALIFA galaxies.  This investigation is useful to reveal what kind of galaxies are mainly involved in the dynamical coherence.

We first divide the CALIFA galaxies into bright ones and faint ones, as shown in Figure~\ref{vprofmagbs}. The criterion magnitude of $M_r=-21.39$ mag is the median value among the CALIFA galaxies.
In this comparison, the trends for central rotation and outskirt rotation are similar to each other in the subsample of bright CALIFA galaxies: both of them hardly show significant signals of coherence. On the other hand, in the subsample of faint CALIFA galaxies, the outskirt rotations show significant (up to $\sim3.5\sigma_{\textrm{\tiny BST}}$ in rel-L weighting) coherence signals, while the central rotations do not (up to $\sim2.0\sigma_{\textrm{\tiny BST}}$ in rel-L weighting), as summarized in Table~\ref{signals2}. 
Like the results from the full CALIFA sample, the results from the luminosity-divided subsamples are more obvious when the rel-L weighting is applied.

Figure~\ref{vprofcolbs} compares the subsamples divided by $g-r$ color. The criterion color of $g-r=0.756$ is the median value among the CALIFA galaxies.
The blue CALIFA galaxies show more significant coherence signals (up to  $\sim3.2\sigma_{\textrm{\tiny BST}}$ in rel-L weighting) than the red CALIFA galaxies, as listed in Table~\ref{signals3}.
However, the coherence signals for the blue CALIFA galaxies are slightly weaker than those for the faint CALIFA galaxies. 

Figure~\ref{vprofserbs} compares the subsamples divided by S{\'e}rsic index. We simply divide the CALIFA galaxies into concentrated ($n\ge 2$) and diffuse ($n<2$) galaxies. In this division, any significant difference between the two subsamples is hardly found. 
One notable feature is that the outskirt rotations of diffuse CALIFA galaxies seem to be very significantly coherent with neighbor motions at $D\leq 30$ kpc ($\sim4.0\sigma_{\textrm{\tiny BST}}$ in rel-L weighting). However, since the significance is low when the RAX uncertainty is used ($\sim1.7\sigma_{\textrm{\tiny RAX}}$), such apparently very high significance to the bootstrap uncertainty may be due to the small number of neighbors within a very close distance of 30 kpc (8 neighbors for the outskirt rotations of the diffuse CALIFA galaxies).
The coherence signals for the subsamples divided by S{\'e}rsic index are summarized in Table~\ref{signals4}.

Finally, in Figure~\ref{vprofmisbs}, we divide the sample by their internal kinematic misalignment, into well-aligned and misaligned galaxies. The division criteria of $|\theta(R\le R_e)-\theta(R_e<R\le 2R_e)|=5.0^{\circ}$ is the median value among the 445 CALIFA galaxies, where $\theta$ is the kinematic position angle at a given radial range.
In the well-aligned CALIFA galaxies, coherence signals are hardly found for both of the central and outskirt angular momenta; the central-outskirt difference is negligible, because they are \emph{well aligned}.
On the other hand, coherence signals are more obvious for the outskirt rotations of the misaligned CALIFA galaxies (up to $\sim2.8\sigma_{\textrm{\tiny BST}}$ in rel-L weighting), as listed in Table~\ref{signals5}.

We conduct more tests for additional constraints on the origin of the dynamical coherence: the division of neighbor galaxies according to their luminosity and color.
In Figure~\ref{neimag}, the right-left-merged mean velocity profiles are shown for our full sample of the CALIFA galaxies, but with the neighbor galaxies controlled by their luminosity. For the rel-L-weighted profiles, the neighbors are divided into `brighter than the adjacent CALIFA galaxy' and `fainter than the adjacent CALIFA galaxy', while they are divided simply by $M_r=-21$ mag cut for the abs-L-weighted profiles. This neighbor division produces a clear difference, in the context that bright neighbors tend to show stronger dynamical coherence (up to 2.6$\sigma_{\textrm{\tiny BST}}$), while faint neighbors hardly show dynamical coherence (up to 1.4$\sigma_{\textrm{\tiny BST}}$).
Figure~\ref{neicol} presents the results for neighbor division into red and blue ones by $g-r=0.756$. In this figure, however, the division into red and blue neighbors produces only small differences. When we consider the color-magnitude relation of neighbor galaxies, such differences between red and blue neighbors may be substantially influenced by the differences between bright and faint neighbors, because fainter galaxies tend to be bluer on average.
The coherence signals in Figures~\ref{neimag} and \ref{neicol} are listed in Tables~\ref{nsig1} and \ref{nsig2}, respectively.

Table~\ref{signalsum} summarizes the overall results for the whole sample and the various subsamples. 
In Table~\ref{signalsum}, the significance to bootstrap uncertainty of coherence signal for each case is simply noted as four classes: significant ($\ge3.0\sigma$), marginal ($2.5 - 2.9\sigma$), very marginal ($2.0 - 2.4\sigma$) and insignificant ($\le 1.9\sigma$), which is convenient for one-shot comparison between the various subsamples. In this simplified summary, the galaxies that are mainly involved in the dynamical coherence appear to be faint, blue or internally-misaligned ones, while the influence of morphology is not clear at least in our sample. When the neighbor galaxies are controlled, bright neighbors appear to influence the outskirt rotations of the CALIFA galaxies more significantly, whereas faint neighbors seem to hardly influence them. The color division of neighbors do not produce notable difference.

\section{DISCUSSION}

\subsection{Confirmation of Dynamical Coherence}

In Section~\ref{result1}, galaxy rotation appears to be related with the average line-of-sight motion of neighbor galaxies. The signals of such dynamical coherence are found up to 800 kpc distance from the CALIFA galaxies in Figure~\ref{vprofbs}. In the range farther than 800 kpc, the coherence signal is insignificant.

This coherence may be the evidence that the interaction of a neighbor galaxy affects the rotational direction of the target galaxy, by adding new angular momentum to it coherent with the flyby direction. This is consistent with the numerical anticipation of \citet{lee18a} that galaxy interactions may affect galaxy rotation as galaxy mergers do, although such changes often cancel out previous changes in their simulations.
However, while the close distance range (e.g., $D\leq$ 200 kpc) is sufficiently close to cause such direct interactions between galaxies, the 800 kpc distance may be too far.\footnote{For example, \citet{woo07} defined `close pairs' as two galaxies within $\Delta v< 500$ {\kms} and $D<50 \,h^{-1}$ kpc.} Thus, those neighbor galaxies may have \emph{recently} experienced interactions with the target galaxies rather than be currently interacting. 

To roughly estimate the timescale that has passed since such interactions, we suppose that the neighbor galaxies at $D\sim800$ kpc experienced flyby events with the target galaxies ($D\sim0$ kpc) and have constantly moved with projected velocity of 180 {\kms}, which is consistent with the peak line-of-sight velocity of the neighbor galaxies at $D\sim0$ kpc in Figure~\ref{vprofbs}(b). The speed of 180 {\kms} corresponds to 184 kpc per Gyr.
Thus, the timescale that a neighbor galaxy travels from $D\sim0$ to 800 kpc is calculated to be $\tau_{\textrm{\tiny T}}\sim4.3$ Gyr.
That is, under our assumption, the vestige of galaxy interactions up to 4 Gyr ago remains as the dynamical coherence with neighbor motions. However, this timescale may change according to the genuine proper motions of the neighbor galaxies that cannot be directly measured.

One of the important results is that the rotation measured at galaxy outskirt ($R_e<R\le 2R_e$) shows stronger dependence on neighbor motions than the rotation measured at galaxy center ($R\le R_e$) does. That is, if it is true that galaxy interactions cause the dynamical coherence discovered in this paper, the effect of such interactions is stronger on the outskirt rotation of the target galaxies rather than on their central rotation.
Many recent studies support the inside-out two-phase formation scenario, in which a massive galaxy formed its central body long time ago by large starburst of major merging events, while its outer body have steadily formed through minor mergers until today \citep[e.g.,][]{ose10,van10,lee13}. Our result is consistent with this scenario in the aspect of galaxy kinematics: recent hierarchical events have strongly influenced the outskirt of a galaxy, whereas their influence on the central body of the galaxy is relatively weak.

\subsection{Implications from the Subsamples}

The investigation of various subsamples provides us useful insights about what kinds of galaxies are mainly involved in the dynamical coherence. In Table~\ref{signalsum}, the most conspicuous result is that the dynamical coherence seems to be mainly related to faint, blue, or internally-misaligned galaxies, particularly for the outskirt rotations.

The dominance of faint galaxies in the dynamical coherence is easy to understand. The kinematic structure of a more massive galaxy may be more difficult to be changed by interactions with its neighbors, because the internal kinematics of a massive galaxy is tightly maintained by its own deep gravitational potential. On the other hand, in the case of a low-mass galaxy, it may be more vulnerable to the influence of its neighbors, because of its shallow gravitational potential. Thus, the internal kinematics of low-mass ($\approx$ faint) galaxies may be more easily influenced by interactions with neighbor galaxies.

Compared to the conspicuous dominance of faint galaxies, the influence of galaxy color seems to be slightly weaker in its statistical significance. Such relative weakness is more obviously revealed in the comparison of the significance to RAX uncertainty. Anyway, in our results, the blue CALIFA galaxies appear to show stronger coherence signals than the red CALIFA galaxies.
If this result reflects some intrinsic difference between the red and blue galaxies, it may indicate that galaxy interactions influence the stellar populations of those galaxies as well as their outskirt rotations. For example, galaxy interactions in moderate distances may trigger more star formation possibly by exchanging their cold gas and giving appropriate perturbation \citep[hydrodynamic interactions;][]{par09}.
However, caution is required for such interpretation, because galaxy color strongly depends on galaxy luminosity. That is, the results for color-divided subsamples may be largely influenced by the fact that most of faint CALIFA galaxies are blue, as shown in Figure~\ref{cmr}.

It is also found that the dynamical coherence of internally-misaligned CALIFA galaxies are stronger than well-aligned CALIFA galaxies.
This result implies that galaxy interactions are largely responsible for the internal kinematic misalignment of galaxies. The rotations of the misaligned galaxies seem to be considerably affected by recent events of galaxy interaction, and such environmental influence is stronger for outskirt rotations than for central rotations, which naturally results in internal kinematic misalignment.
In other words, galaxy interactions, as well as major mergers \citep{boi11} or minor mergers \citep{tay18}, may cause the internal misalignment in a galaxy, the extreme cases of which may be KDCs.

In summary, the dynamical coherence is mainly due to faint, blue or internally-misaligned galaxies. This seems to be because low-mass galaxies are easier to be kinematically affected by galaxy interactions, the influence of which may be differential according to radius. The change of stellar populations during such interactions may be possible, but it is too rash to conclude about it here.

\subsection{Interacted? Or Inherited?}

One may suspect if the observed dynamical coherence is inherited from the in-situ rotation of host dark matter halos or past shared gas streams. If a central galaxy and its satellite galaxies formed in a single rotating halo, then their co-rotation may be natural.
Moreover, high-redshift galaxies are often expected to be formed by collimated streams of material that transfer angular momentum to the central galaxy. Such streams may strongly affect the rotation of the central galaxy and the motions of its satellites at the same time, and thus may cause the dynamical coherence.
We check how well this scenario is supported by our results.

If the dynamical coherence is the results from the in-situ co-rotation of a central galaxy and its satellite galaxies, then it is expected that our target CALIFA galaxies tend to be central galaxies in small groups and the co-rotating neighbors are their low-mass satellites. However, our results show the opposite trends: faint CALIFA galaxies show stronger dynamical coherence than bright CALIFA galaxies that are more probable to be central galaxies. Moreover, in the test dividing neighbors by luminosity, faint neighbors hardly show any coherence signals regardless of central and outskirt rotations in rel-L and abs-L weightings. On the other hand, bright neighbors have strong coherence signals for the outskirt rotations of the CALIFA galxies. Such signals are stronger when the rel-L weighting is applied than when the abs-L weighting is applied, which implies that more crucial for dynamical coherence is `how \emph{more} massive the neighbor is' than `how massive the neighbor is'. Obviously, all of these results support the interaction origin rather than the inherited co-rotation.

In summary, our results indicate that the main origin of the observed coherence between galaxy rotations and neighbor motions may be their recent interactions. Although this does not completely deny the partial contribution of in-situ co-rotations, we hardly find any evidence clearly in favor of the inherited co-rotation scenario in our current results.

\section{CONCLUSION}

We examined whether there is any coherence between the rotational direction of galaxies and the average motion of their neighbor galaxies, using the CALIFA survey data and the NSA catalog.
From our statistical analysis, we discovered that such coherence actually exists.
Our main conclusions are summarized as follows:
\begin{enumerate}
 \item[1.] The rotation of a galaxy appears to be significantly influenced by the average motion of its neighbor galaxies (up to $\sim800$ kpc). Recent events of galaxy interactions (possibly up to 4 Gyr ago) may have influenced galaxy rotation.
 \item[2.] The dynamical coherence is stronger for outskirt rotations ($R_e<R\le 2R_e$) than central rotations ($R\le R_e$). This result is consistent with the inside-out two-phase formation scenario of massive galaxies, in the context that recent hierarchical events may have mainly influenced outer body of a galaxy rather than its central body.
 \item[3.] Faint galaxies tend to show more conspicuous signals of dynamical coherence than bright galaxies do. Furthermore, the coherence signals are much stronger for bright neighbors. These indicate that the internal kinematics of more massive galaxies are less affected by interactions with neighbors, while more massive neighbors tend to influence galaxy rotation more strongly.
 \item[4.] Internally misaligned galaxies tend to show more conspicuous signals of dynamical coherence than internally well-aligned galaxies do. That is, interactions with nearby neighbors seem to be largely responsible for the internal kinematic misalignment of a galaxy or even a kinematically distinct core.
\end{enumerate}

In this paper, we presented not only the discovery of observational evidence for the coherence between galaxy rotation and neighbor motions, but also how such a trend depends on the properties of target galaxies. Although our sample size is enough to prove the existence itself of the dynamical coherence, it is not enough for confirming some details of various subsamples, because the statistical uncertainty is often too large when the sample is finely divided. For example, if we secure a sufficient number of IFS data for faint-but-red galaxies, we can better constrain the relationship between the dynamical coherence and stellar populations. This limit is expected to be improved in near future, if data from sufficiently large IFS surveys with large field-of-view (covering at least $2R_e$ of target galaxies) become available. Otherwise, the Mapping Nearby Galaxies at Apache Point Observatory \citep[MaNGA;][]{bun15} data may be an alternative: although its basic spatial coverage for each target is only $1.5 R_e$, its large sample size (up to 10,000 nearby galaxies) will be a great merit to enhance statistical reliability.

\begin{deluxetable*}{ccrrrrrrr}
\rotate
\tablenum{2} \tablecolumns{9} \tablecaption{Coherence Signals (CALIFA Galaxies Divided by Luminosity)} \tablewidth{0pt}
\tablehead{ & & & \multicolumn{3}{c}{Bright CALIFA Galaxies} & \multicolumn{3}{c}{Faint CALIFA Galaxies} \\
Weighting & Angular & $D$ & $\langle\Delta v\rangle^{c}_{R-L} \pm\sigma_{\textrm{\tiny BST}}$ & $\pm\sigma_{\textrm{\tiny RAX}}$ & $\pm\sigma_{\textrm{\tiny RFA}}$ & $\langle\Delta v\rangle^{c}_{R-L} \pm\sigma_{\textrm{\tiny BST}}$ & $\pm\sigma_{\textrm{\tiny RAX}}$ & $\pm\sigma_{\textrm{\tiny RFA}}$\\
& momentum & [kpc] & [km s$^{-1}$] & [km s$^{-1}$] & [km s$^{-1}$] & [km s$^{-1}$] & [km s$^{-1}$] & [km s$^{-1}$]}
\startdata
Rel-L & $R\leq R_e$ & 30 & $64.0\pm252.7\,(0.3\sigma)$ & $\pm135.3\,(0.5\sigma)$ & $\pm152.7\,(0.4\sigma)$ & $67.0\pm76.5\,(0.9\sigma)$ & $\pm58.2\,(1.2\sigma)$ & $\pm77.5\,(0.9\sigma)$ \\ 
 & & 200 & $13.5\pm62.4\,(0.2\sigma)$ & $\pm51.0\,(0.3\sigma)$ & $\pm51.8\,(0.3\sigma)$ & $60.1\pm47.6\,(1.3\sigma)$ & $\pm39.8\,(1.5\sigma)$ & $\pm37.4\,(1.6\sigma)$ \\ 
 & & 500 & $10.0\pm32.7\,(0.3\sigma)$ & $\pm25.0\,(0.4\sigma)$ & $\pm22.7\,(0.4\sigma)$ & $26.3\pm25.3\,(1.0\sigma)$ & $\pm22.8\,(1.2\sigma)$ & $\pm24.3\,(1.1\sigma)$ \\ 
 & & 800 & $-30.4\pm21.5\,(1.4\sigma)$ & $\pm19.0\,(1.6\sigma)$ & $\pm16.1\,(1.9\sigma)$ & $35.4\pm17.3\,(2.0\sigma)$ & $\pm19.8\,(1.8\sigma)$ & $\pm17.4\,(2.0\sigma)$ \\ 
\hline
Rel-L & $R_e<R\leq 2R_e$ & 30 & $122.6\pm213.8\,(0.6\sigma)$ & $\pm141.4\,(0.9\sigma)$ & $\pm163.3\,(0.8\sigma)$ & $149.3\pm53.8\,(2.8\sigma)$ & $\pm78.2\,(1.9\sigma)$ & $\pm52.5\,(2.8\sigma)$ \\  
 & & 200 & $23.2\pm60.2\,(0.4\sigma)$ & $\pm52.5\,(0.4\sigma)$ & $\pm66.4\,(0.3\sigma)$ & $114.8\pm64.4\,(1.8\sigma)$ & $\pm53.6\,(2.1\sigma)$ & $\pm35.7\,(3.2\sigma)$ \\ 
 & & 500 & $8.2\pm36.2\,(0.2\sigma)$ & $\pm25.9\,(0.3\sigma)$ & $\pm26.4\,(0.3\sigma)$ & $65.8\pm31.4\,(2.1\sigma)$ & $\pm27.9\,(2.4\sigma)$ & $\pm24.3\,(2.7\sigma)$ \\
 & & 800 & $-35.2\pm22.4\,(1.6\sigma)$ & $\pm20.6\,(1.7\sigma)$ & $\pm17.1\,(2.1\sigma)$ & $70.6\pm20.0\,(3.5\sigma)$ & $\pm22.5\,(3.1\sigma)$ & $\pm16.9\,(4.2\sigma)$ \\ 
\hline
Abs-L & $R\leq R_e$ & 30 & $39.0\pm250.2\,(0.2\sigma)$ & $\pm142.4\,(0.3\sigma)$ & $\pm174.8\,(0.2\sigma)$ & $89.1\pm75.9\,(1.2\sigma)$ & $\pm65.1\,(1.4\sigma)$ & $\pm73.2\,(1.2\sigma)$ \\ 
 & & 200 & $21.5\pm58.4\,(0.4\sigma)$ & $\pm47.4\,(0.5\sigma)$ & $\pm49.4\,(0.4\sigma)$ & $75.9\pm43.2\,(1.8\sigma)$ & $\pm32.5\,(2.3\sigma)$ & $\pm26.3\,(2.9\sigma)$ \\ 
 & & 500 & $16.8\pm28.4\,(0.6\sigma)$ & $\pm22.1\,(0.8\sigma)$ & $\pm21.1\,(0.8\sigma)$ & $39.1\pm26.2\,(1.5\sigma)$ & $\pm24.8\,(1.6\sigma)$ & $\pm24.7\,(1.6\sigma)$ \\ 
 & & 800 & $-27.7\pm21.0\,(1.3\sigma)$ & $\pm18.1\,(1.5\sigma)$ & $\pm13.7\,(2.0\sigma)$ & $30.3\pm16.7\,(1.8\sigma)$ & $\pm18.7\,(1.6\sigma)$ & $\pm19.2\,(1.6\sigma)$ \\ 
\hline
Abs-L & $R_e<R\leq 2R_e$ & 30 & $120.3\pm223.7\,(0.5\sigma)$ & $\pm145.3\,(0.8\sigma)$ & $\pm180.8\,(0.7\sigma)$ & $138.7\pm57.4\,(2.4\sigma)$ & $\pm68.2\,(2.0\sigma)$ & $\pm58.5\,(2.4\sigma)$ \\ 
 & & 200 & $37.3\pm56.9\,(0.7\sigma)$ & $\pm51.1\,(0.7\sigma)$ & $\pm61.5\,(0.6\sigma)$ & $85.7\pm46.2\,(1.9\sigma)$ & $\pm31.3\,(2.7\sigma)$ & $\pm26.8\,(3.2\sigma)$ \\ 
 & & 500 & $14.7\pm30.6\,(0.5\sigma)$ & $\pm23.0\,(0.6\sigma)$ & $\pm21.6\,(0.7\sigma)$ & $55.1\pm28.8\,(1.9\sigma)$ & $\pm24.5\,(2.2\sigma)$ & $\pm23.8\,(2.3\sigma)$ \\ 
 & & 800 & $-35.9\pm21.8\,(1.7\sigma)$ & $\pm18.2\,(2.0\sigma)$ & $\pm14.3\,(2.5\sigma)$ & $48.0\pm17.9\,(2.7\sigma)$ & $\pm20.0\,(2.4\sigma)$ & $\pm17.2\,(2.8\sigma)$
\enddata
\tablecomments{}
\label{signals2}
\end{deluxetable*}

\begin{deluxetable*}{ccrrrrrrr}
\rotate
\tablenum{3} \tablecolumns{9} \tablecaption{Coherence Signals (CALIFA Galaxies Divided by Color)} \tablewidth{0pt}
\tablehead{ & & & \multicolumn{3}{c}{Red CALIFA Galaxies} & \multicolumn{3}{c}{Blue CALIFA Galaxies} \\
Weighting & Angular & $D$ & $\langle\Delta v\rangle^{c}_{R-L} \pm\sigma_{\textrm{\tiny BST}}$ & $\pm\sigma_{\textrm{\tiny RAX}}$ & $\pm\sigma_{\textrm{\tiny RFA}}$ & $\langle\Delta v\rangle^{c}_{R-L} \pm\sigma_{\textrm{\tiny BST}}$ & $\pm\sigma_{\textrm{\tiny RAX}}$ & $\pm\sigma_{\textrm{\tiny RFA}}$\\
& momentum & [kpc] & [km s$^{-1}$] & [km s$^{-1}$] & [km s$^{-1}$] & [km s$^{-1}$] & [km s$^{-1}$] & [km s$^{-1}$]}
\startdata
Rel-L & $R\leq R_e$ & 30 & $95.6\pm105.2\,(0.9\sigma)$ & $\pm95.1\,(1.0\sigma)$ & $\pm136.6\,(0.7\sigma)$ & $53.0\pm126.3\,(0.4\sigma)$ & $\pm42.3\,(1.3\sigma)$ & $\pm39.5\,(1.3\sigma)$ \\ 
 & & 200 & $51.7\pm70.5\,(0.7\sigma)$ & $\pm46.7\,(1.1\sigma)$ & $\pm52.4\,(1.0\sigma)$ & $59.4\pm57.0\,(1.0\sigma)$ & $\pm46.7\,(1.3\sigma)$ & $\pm40.1\,(1.5\sigma)$ \\ 
 & & 500 & $27.0\pm35.1\,(0.8\sigma)$ & $\pm29.4\,(0.9\sigma)$ & $\pm30.2\,(0.9\sigma)$ & $24.6\pm30.8\,(0.8\sigma)$ & $\pm27.9\,(0.9\sigma)$ & $\pm29.5\,(0.8\sigma)$ \\ 
 & & 800 & $21.0\pm21.5\,(1.0\sigma)$ & $\pm24.4\,(0.9\sigma)$ & $\pm23.8\,(0.9\sigma)$ & $34.5\pm21.1\,(1.6\sigma)$ & $\pm23.9\,(1.4\sigma)$ & $\pm19.5\,(1.8\sigma)$ \\  
\hline
Rel-L & $R_e<R\leq 2R_e$ & 30 & $174.7\pm95.0\,(1.8\sigma)$ & $\pm97.5\,(1.8\sigma)$ & $\pm88.5\,(2.0\sigma)$ & $98.3\pm44.9\,(2.2\sigma)$ & $\pm66.8\,(1.5\sigma)$ & $\pm52.8\,(1.9\sigma)$ \\  
 & & 200 & $77.6\pm61.6\,(1.3\sigma)$ & $\pm49.4\,(1.6\sigma)$ & $\pm53.3\,(1.5\sigma)$ & $121.1\pm77.8\,(1.6\sigma)$ & $\pm71.0\,(1.7\sigma)$ & $\pm37.6\,(3.2\sigma)$ \\ 
 & & 500 & $29.5\pm36.5\,(0.8\sigma)$ & $\pm31.1\,(0.9\sigma)$ & $\pm26.0\,(1.1\sigma)$ & $74.6\pm37.0\,(2.0\sigma)$ & $\pm36.0\,(2.1\sigma)$ & $\pm34.1\,(2.2\sigma)$ \\
 & & 800 & $31.8\pm23.8\,(1.3\sigma)$ & $\pm25.8\,(1.2\sigma)$ & $\pm18.9\,(1.7\sigma)$ & $74.6\pm23.5\,(3.2\sigma)$ & $\pm28.5\,(2.6\sigma)$ & $\pm22.9\,(3.3\sigma)$ \\  
\hline
Abs-L & $R\leq R_e$ & 30 & $59.5\pm140.6\,(0.4\sigma)$ & $\pm89.2\,(0.7\sigma)$ & $\pm99.7\,(0.6\sigma)$ & $22.5\pm137.7\,(0.2\sigma)$ & $\pm49.2\,(0.5\sigma)$ & $\pm55.6\,(0.4\sigma)$ \\  
 & & 200 & $37.8\pm45.1\,(0.8\sigma)$ & $\pm32.8\,(1.2\sigma)$ & $\pm32.1\,(1.2\sigma)$ & $87.5\pm55.2\,(1.6\sigma)$ & $\pm47.4\,(1.8\sigma)$ & $\pm39.4\,(2.2\sigma)$ \\ 
 & & 500 & $31.3\pm25.7\,(1.2\sigma)$ & $\pm20.0\,(1.6\sigma)$ & $\pm19.9\,(1.6\sigma)$ & $28.0\pm31.4\,(0.9\sigma)$ & $\pm27.9\,(1.0\sigma)$ & $\pm29.1\,(1.0\sigma)$ \\ 
 & & 800 & $3.4\pm18.2\,(0.2\sigma)$ & $\pm15.9\,(0.2\sigma)$ & $\pm15.2\,(0.2\sigma)$ & $16.0\pm18.5\,(0.9\sigma)$ & $\pm20.2\,(0.8\sigma)$ & $\pm20.7\,(0.8\sigma)$ \\ 
\hline
Abs-L & $R_e<R\leq 2R_e$ & 30 & $137.3\pm115.4\,(1.2\sigma)$ & $\pm95.2\,(1.4\sigma)$ & $\pm127.0\,(1.1\sigma)$ & $82.5\pm41.4\,(2.0\sigma)$ & $\pm68.0\,(1.2\sigma)$ & $\pm83.5\,(1.0\sigma)$ \\ 
 & & 200 & $57.4\pm47.3\,(1.2\sigma)$ & $\pm34.4\,(1.7\sigma)$ & $\pm35.7\,(1.6\sigma)$ & $91.0\pm54.4\,(1.7\sigma)$ & $\pm47.0\,(1.9\sigma)$ & $\pm43.9\,(2.1\sigma)$ \\ 
 & & 500 & $25.4\pm27.3\,(0.9\sigma)$ & $\pm21.6\,(1.2\sigma)$ & $\pm20.4\,(1.2\sigma)$ & $63.7\pm32.4\,(2.0\sigma)$ & $\pm29.6\,(2.2\sigma)$ & $\pm24.9\,(2.6\sigma)$ \\ 
 & & 800 & $-0.3\pm18.5\,(0.0\sigma)$ & $\pm17.6\,(0.0\sigma)$ & $\pm14.3\,(0.0\sigma)$ & $36.8\pm21.5\,(1.7\sigma)$ & $\pm22.5\,(1.6\sigma)$ & $\pm17.3\,(2.1\sigma)$
\enddata
\tablecomments{}
\label{signals3}
\end{deluxetable*}

\begin{deluxetable*}{ccrrrrrrr}
\rotate
\tablenum{4} \tablecolumns{9} \tablecaption{Coherence Signals (CALIFA Galaxies Divided by S{\'e}rsic Index)} \tablewidth{0pt}
\tablehead{ & & & \multicolumn{3}{c}{Concentrated CALIFA Galaxies} & \multicolumn{3}{c}{Diffuse CALIFA Galaxies} \\
Weighting & Angular & $D$ & $\langle\Delta v\rangle^{c}_{R-L} \pm\sigma_{\textrm{\tiny BST}}$ & $\pm\sigma_{\textrm{\tiny RAX}}$ & $\pm\sigma_{\textrm{\tiny RFA}}$ & $\langle\Delta v\rangle^{c}_{R-L} \pm\sigma_{\textrm{\tiny BST}}$ & $\pm\sigma_{\textrm{\tiny RAX}}$ & $\pm\sigma_{\textrm{\tiny RFA}}$\\
& momentum & [kpc] & [km s$^{-1}$] & [km s$^{-1}$] & [km s$^{-1}$] & [km s$^{-1}$] & [km s$^{-1}$] & [km s$^{-1}$]}
\startdata
Rel-L & $R\leq R_e$ & 30 & $82.8\pm112.4\,(0.7\sigma)$ & $\pm53.5\,(1.5\sigma)$ & $\pm46.7\,(1.8\sigma)$ & $58.6\pm44.2\,(1.3\sigma)$ & $\pm79.6\,(0.7\sigma)$ & $\pm104.2\,(0.6\sigma)$ \\  
 & & 200 & $86.9\pm71.7\,(1.2\sigma)$ & $\pm57.9\,(1.5\sigma)$ & $\pm27.5\,(3.2\sigma)$ & $27.2\pm54.7\,(0.5\sigma)$ & $\pm37.9\,(0.7\sigma)$ & $\pm54.2\,(0.5\sigma)$ \\ 
 & & 500 & $37.1\pm37.4\,(1.0\sigma)$ & $\pm33.8\,(1.1\sigma)$ & $\pm27.2\,(1.4\sigma)$ & $17.0\pm28.7\,(0.6\sigma)$ & $\pm27.1\,(0.6\sigma)$ & $\pm31.3\,(0.5\sigma)$ \\ 
 & & 800 & $32.7\pm22.1\,(1.5\sigma)$ & $\pm26.8\,(1.2\sigma)$ & $\pm16.4\,(2.0\sigma)$ & $30.3\pm20.6\,(1.5\sigma)$ & $\pm24.4\,(1.2\sigma)$ & $\pm21.9\,(1.4\sigma)$ \\  
\hline
Rel-L & $R_e<R\leq 2R_e$ & 30 & $121.2\pm92.3\,(1.3\sigma)$ & $\pm68.2\,(1.8\sigma)$ & $\pm63.9\,(1.9\sigma)$ & $185.8\pm46.9\,(4.0\sigma)$ & $\pm109.8\,(1.7\sigma)$ & $\pm68.5\,(2.7\sigma)$ \\
 & & 200 & $105.7\pm73.4\,(1.4\sigma)$ & $\pm58.9\,(1.8\sigma)$ & $\pm27.0\,(3.9\sigma)$ & $115.5\pm81.4\,(1.4\sigma)$ & $\pm74.5\,(1.5\sigma)$ & $\pm62.9\,(1.8\sigma)$ \\
 & & 500 & $70.5\pm38.6\,(1.8\sigma)$ & $\pm34.6\,(2.0\sigma)$ & $\pm27.5\,(2.6\sigma)$ & $53.7\pm36.8\,(1.5\sigma)$ & $\pm37.3\,(1.4\sigma)$ & $\pm33.8\,(1.6\sigma)$ \\
 & & 800 & $57.4\pm24.1\,(2.4\sigma)$ & $\pm28.2\,(2.0\sigma)$ & $\pm15.1\,(3.8\sigma)$ & $64.7\pm23.4\,(2.8\sigma)$ & $\pm27.8\,(2.3\sigma)$ & $\pm24.3\,(2.7\sigma)$ \\  
\hline
Abs-L & $R\leq R_e$ & 30 & $57.1\pm139.6\,(0.4\sigma)$ & $\pm92.6\,(0.6\sigma)$ & $\pm106.3\,(0.5\sigma)$ & $41.4\pm42.5\,(1.0\sigma)$ & $\pm105.9\,(0.4\sigma)$ & $\pm120.0\,(0.3\sigma)$ \\
 & & 200 & $51.3\pm42.3\,(1.2\sigma)$ & $\pm30.8\,(1.7\sigma)$ & $\pm29.8\,(1.7\sigma)$ & $77.2\pm64.1\,(1.2\sigma)$ & $\pm52.1\,(1.5\sigma)$ & $\pm43.2\,(1.8\sigma)$ \\ 
 & & 500 & $25.4\pm23.9\,(1.1\sigma)$ & $\pm19.9\,(1.3\sigma)$ & $\pm19.6\,(1.3\sigma)$ & $45.3\pm34.1\,(1.3\sigma)$ & $\pm30.8\,(1.5\sigma)$ & $\pm28.4\,(1.6\sigma)$ \\ 
 & & 800 & $4.3\pm16.3\,(0.3\sigma)$ & $\pm16.1\,(0.3\sigma)$ & $\pm13.7\,(0.3\sigma)$ & $18.3\pm20.3\,(0.9\sigma)$ & $\pm22.4\,(0.8\sigma)$ & $\pm22.9\,(0.8\sigma)$ \\ 
\hline
Abs-L & $R_e<R\leq 2R_e$ & 30 & $121.7\pm108.1\,(1.1\sigma)$ & $\pm92.7\,(1.3\sigma)$ & $\pm131.6\,(0.9\sigma)$ & $162.9\pm43.8\,(3.7\sigma)$ & $\pm119.7\,(1.4\sigma)$ & $\pm93.6\,(1.7\sigma)$ \\
 & & 200 & $63.0\pm41.9\,(1.5\sigma)$ & $\pm32.2\,(2.0\sigma)$ & $\pm34.7\,(1.8\sigma)$ & $92.0\pm63.6\,(1.4\sigma)$ & $\pm51.5\,(1.8\sigma)$ & $\pm46.3\,(2.0\sigma)$ \\ 
 & & 500 & $33.3\pm24.7\,(1.3\sigma)$ & $\pm20.0\,(1.7\sigma)$ & $\pm19.4\,(1.7\sigma)$ & $55.9\pm38.1\,(1.5\sigma)$ & $\pm33.5\,(1.7\sigma)$ & $\pm25.7\,(2.2\sigma)$ \\ 
 & & 800 & $5.6\pm17.5\,(0.3\sigma)$ & $\pm15.9\,(0.4\sigma)$ & $\pm13.6\,(0.4\sigma)$ & $33.9\pm22.9\,(1.5\sigma)$ & $\pm25.4\,(1.3\sigma)$ & $\pm20.5\,(1.7\sigma)$
\enddata
\tablecomments{}
\label{signals4}
\end{deluxetable*}

\begin{deluxetable*}{ccrrrrrrr}
\rotate
\tablenum{5} \tablecolumns{9} \tablecaption{Coherence Signals (CALIFA Galaxies Divided by Misalignment)} \tablewidth{0pt}
\tablehead{ & & & \multicolumn{3}{c}{Aligned CALIFA Galaxies} & \multicolumn{3}{c}{Misaligned CALIFA Galaxies} \\
Weighting & Angular & $D$ & $\langle\Delta v\rangle^{c}_{R-L} \pm\sigma_{\textrm{\tiny BST}}$ & $\pm\sigma_{\textrm{\tiny RAX}}$ & $\pm\sigma_{\textrm{\tiny RFA}}$ & $\langle\Delta v\rangle^{c}_{R-L} \pm\sigma_{\textrm{\tiny BST}}$ & $\pm\sigma_{\textrm{\tiny RAX}}$ & $\pm\sigma_{\textrm{\tiny RFA}}$\\
& momentum & [kpc] & [km s$^{-1}$] & [km s$^{-1}$] & [km s$^{-1}$] & [km s$^{-1}$] & [km s$^{-1}$] & [km s$^{-1}$]}
\startdata
Rel-L & $R\leq R_e$ & 30 & $77.5\pm83.9\,(0.9\sigma)$ & $\pm91.2\,(0.8\sigma)$ & $\pm118.6\,(0.7\sigma)$ & $100.0\pm122.1\,(0.8\sigma)$ & $\pm89.0\,(1.1\sigma)$ & $\pm105.4\,(0.9\sigma)$ \\
 & & 200 & $6.5\pm56.0\,(0.1\sigma)$ & $\pm41.6\,(0.2\sigma)$ & $\pm30.1\,(0.2\sigma)$ & $86.3\pm75.4\,(1.1\sigma)$ & $\pm58.1\,(1.5\sigma)$ & $\pm52.9\,(1.6\sigma)$ \\  
 & & 500 & $-2.3\pm37.2\,(0.1\sigma)$ & $\pm36.3\,(0.1\sigma)$ & $\pm44.2\,(0.1\sigma)$ & $40.9\pm36.3\,(1.1\sigma)$ & $\pm33.6\,(1.2\sigma)$ & $\pm30.9\,(1.3\sigma)$ \\
 & & 800 & $-7.2\pm24.6\,(0.3\sigma)$ & $\pm27.8\,(0.3\sigma)$ & $\pm33.6\,(0.2\sigma)$ & $48.2\pm23.6\,(2.0\sigma)$ & $\pm31.4\,(1.5\sigma)$ & $\pm25.0\,(1.9\sigma)$ \\
\hline
Rel-L & $R_e<R\leq 2R_e$ & 30 & $78.0\pm83.7\,(0.9\sigma)$ & $\pm93.1\,(0.8\sigma)$ & $\pm118.3\,(0.7\sigma)$ & $179.8\pm63.3\,(2.8\sigma)$ & $\pm88.5\,(2.0\sigma)$ & $\pm51.5\,(3.5\sigma)$ \\ 
 & & 200 & $28.1\pm57.5\,(0.5\sigma)$ & $\pm40.4\,(0.7\sigma)$ & $\pm37.8\,(0.7\sigma)$ & $107.1\pm69.2\,(1.5\sigma)$ & $\pm58.1\,(1.8\sigma)$ & $\pm45.0\,(2.4\sigma)$ \\ 
 & & 500 & $32.7\pm37.7\,(0.9\sigma)$ & $\pm37.5\,(0.9\sigma)$ & $\pm37.2\,(0.9\sigma)$ & $56.8\pm38.3\,(1.5\sigma)$ & $\pm33.0\,(1.7\sigma)$ & $\pm27.6\,(2.1\sigma)$ \\
 & & 800 & $18.2\pm25.6\,(0.7\sigma)$ & $\pm28.2\,(0.6\sigma)$ & $\pm27.2\,(0.7\sigma)$ & $66.1\pm23.3\,(2.8\sigma)$ & $\pm25.9\,(2.6\sigma)$ & $\pm18.7\,(3.5\sigma)$ \\  
\hline
Abs-L & $R\leq R_e$ & 30 & $66.8\pm90.0\,(0.7\sigma)$ & $\pm84.9\,(0.8\sigma)$ & $\pm89.4\,(0.7\sigma)$ & $56.3\pm157.5\,(0.4\sigma)$ & $\pm104.4\,(0.5\sigma)$ & $\pm110.1\,(0.5\sigma)$ \\ 
 & & 200 & $21.2\pm64.7\,(0.3\sigma)$ & $\pm45.6\,(0.5\sigma)$ & $\pm43.0\,(0.5\sigma)$ & $62.8\pm43.3\,(1.5\sigma)$ & $\pm34.6\,(1.8\sigma)$ & $\pm35.7\,(1.8\sigma)$ \\ 
 & & 500 & $0.6\pm34.8\,(0.0\sigma)$ & $\pm29.7\,(0.0\sigma)$ & $\pm32.1\,(0.0\sigma)$ & $49.0\pm21.4\,(2.3\sigma)$ & $\pm20.9\,(2.3\sigma)$ & $\pm21.0\,(2.3\sigma)$ \\  
 & & 800 & $-10.9\pm24.0\,(0.5\sigma)$ & $\pm20.0\,(0.5\sigma)$ & $\pm20.0\,(0.5\sigma)$ & $19.5\pm16.3\,(1.2\sigma)$ & $\pm17.5\,(1.1\sigma)$ & $\pm16.6\,(1.2\sigma)$ \\
\hline
Abs-L & $R_e<R\leq 2R_e$ & 30 & $67.4\pm89.7\,(0.8\sigma)$ & $\pm86.8\,(0.8\sigma)$ & $\pm90.6\,(0.7\sigma)$ & $147.4\pm95.2\,(1.5\sigma)$ & $\pm105.8\,(1.4\sigma)$ & $\pm134.9\,(1.1\sigma)$ \\ 
 & & 200 & $25.8\pm65.8\,(0.4\sigma)$ & $\pm45.2\,(0.6\sigma)$ & $\pm45.7\,(0.6\sigma)$ & $86.5\pm44.4\,(1.9\sigma)$ & $\pm34.3\,(2.5\sigma)$ & $\pm38.1\,(2.3\sigma)$ \\ 
 & & 500 & $15.4\pm35.4\,(0.4\sigma)$ & $\pm29.4\,(0.5\sigma)$ & $\pm28.2\,(0.5\sigma)$ & $56.1\pm22.4\,(2.5\sigma)$ & $\pm20.4\,(2.7\sigma)$ & $\pm19.0\,(2.9\sigma)$ \\ 
 & & 800 & $0.2\pm23.8\,(0.0\sigma)$ & $\pm20.9\,(0.0\sigma)$ & $\pm17.6\,(0.0\sigma)$ & $20.3\pm17.1\,(1.2\sigma)$ & $\pm17.6\,(1.2\sigma)$ & $\pm14.6\,(1.4\sigma)$
\enddata
\tablecomments{}
\label{signals5}
\end{deluxetable*}

\begin{deluxetable*}{ccrrrrrrr}
\rotate
\tablenum{6} \tablecolumns{9} \tablecaption{Coherence Signals (Neighbors Divided by Luminosity)} \tablewidth{0pt}
\tablehead{ & & & \multicolumn{3}{c}{Bright Neighbors} & \multicolumn{3}{c}{Faint Neighbors} \\
Weighting & Angular & $D$ & $\langle\Delta v\rangle^{c}_{R-L} \pm\sigma_{\textrm{\tiny BST}}$ & $\pm\sigma_{\textrm{\tiny RAX}}$ & $\pm\sigma_{\textrm{\tiny RFA}}$ & $\langle\Delta v\rangle^{c}_{R-L} \pm\sigma_{\textrm{\tiny BST}}$ & $\pm\sigma_{\textrm{\tiny RAX}}$ & $\pm\sigma_{\textrm{\tiny RFA}}$\\
& momentum & [kpc] & [km s$^{-1}$] & [km s$^{-1}$] & [km s$^{-1}$] & [km s$^{-1}$] & [km s$^{-1}$] & [km s$^{-1}$]}
\startdata
Rel-L & $R\leq R_e$ & 30 & $84.1\pm97.7\,(0.9\sigma)$ & $\pm81.5\,(1.0\sigma)$ & $\pm100.2\,(0.8\sigma)$ & $24.1\pm95.0\,(0.3\sigma)$ & $\pm44.3\,(0.5\sigma)$ & $\pm48.2\,(0.5\sigma)$ \\ 
 & & 200 & $71.6\pm81.2\,(0.9\sigma)$ & $\pm47.7\,(1.5\sigma)$ & $\pm48.9\,(1.5\sigma)$ & $-2.9\pm50.7\,(0.1\sigma)$ & $\pm22.0\,(0.1\sigma)$ & $\pm24.9\,(0.1\sigma)$ \\
 & & 500 & $32.7\pm47.2\,(0.7\sigma)$ & $\pm27.9\,(1.2\sigma)$ & $\pm30.1\,(1.1\sigma)$ & $2.0\pm27.3\,(0.1\sigma)$ & $\pm13.7\,(0.1\sigma)$ & $\pm12.8\,(0.2\sigma)$ \\  
 & & 800 & $39.8\pm34.0\,(1.2\sigma)$ & $\pm23.4\,(1.7\sigma)$ & $\pm19.8\,(2.0\sigma)$ & $6.3\pm18.4\,(0.3\sigma)$ & $\pm11.3\,(0.6\sigma)$ & $\pm11.0\,(0.6\sigma)$ \\  
\hline
Rel-L & $R_e<R\leq 2R_e$ & 30 & $173.7\pm66.1\,(2.6\sigma)$ & $\pm97.6\,(1.8\sigma)$ & $\pm68.0\,(2.6\sigma)$ & $73.5\pm51.1\,(1.4\sigma)$ & $\pm57.5\,(1.3\sigma)$ & $\pm68.3\,(1.1\sigma)$ \\  
 & & 200 & $138.0\pm92.3\,(1.5\sigma)$ & $\pm67.5\,(2.0\sigma)$ & $\pm44.5\,(3.1\sigma)$ & $5.0\pm67.0\,(0.1\sigma)$ & $\pm23.5\,(0.2\sigma)$ & $\pm27.5\,(0.2\sigma)$ \\  
 & & 500 & $83.4\pm56.9\,(1.5\sigma)$ & $\pm34.3\,(2.4\sigma)$ & $\pm29.2\,(2.9\sigma)$ & $0.9\pm30.9\,(0.0\sigma)$ & $\pm13.9\,(0.1\sigma)$ & $\pm14.1\,(0.1\sigma)$ \\ 
 & & 800 & $89.6\pm39.2\,(2.3\sigma)$ & $\pm27.4\,(3.3\sigma)$ & $\pm19.2\,(4.7\sigma)$ & $-0.1\pm19.1\,(0.0\sigma)$ & $\pm11.7\,(0.0\sigma)$ & $\pm10.7\,(0.0\sigma)$ \\
\hline
Abs-L & $R\leq R_e$ & 30 & $81.8\pm182.0\,(0.4\sigma)$ & $\pm138.8\,(0.6\sigma)$ & $\pm169.1\,(0.5\sigma)$ & $-3.4\pm108.8\,(0.0\sigma)$ & $\pm41.2\,(0.1\sigma)$ & $\pm43.4\,(0.1\sigma)$ \\
 & & 200 & $97.1\pm93.0\,(1.0\sigma)$ & $\pm49.9\,(1.9\sigma)$ & $\pm51.5\,(1.9\sigma)$ & $11.0\pm37.9\,(0.3\sigma)$ & $\pm21.7\,(0.5\sigma)$ & $\pm22.4\,(0.5\sigma)$ \\ 
 & & 500 & $56.4\pm52.3\,(1.1\sigma)$ & $\pm31.3\,(1.8\sigma)$ & $\pm30.7\,(1.8\sigma)$ & $10.8\pm21.6\,(0.5\sigma)$ & $\pm13.8\,(0.8\sigma)$ & $\pm15.2\,(0.7\sigma)$ \\ 
 & & 800 & $8.5\pm42.5\,(0.2\sigma)$ & $\pm23.5\,(0.4\sigma)$ & $\pm21.9\,(0.4\sigma)$ & $8.1\pm14.4\,(0.6\sigma)$ & $\pm11.5\,(0.7\sigma)$ & $\pm11.8\,(0.7\sigma)$ \\ 
\hline
Abs-L & $R_e<R\leq 2R_e$ & 30 & $173.7\pm77.7\,(2.2\sigma)$ & $\pm127.5\,(1.4\sigma)$ & $\pm196.7\,(0.9\sigma)$ & $-1.0\pm77.3\,(0.0\sigma)$ & $\pm47.1\,(0.0\sigma)$ & $\pm54.9\,(0.0\sigma)$ \\
 & & 200 & $120.4\pm80.5\,(1.5\sigma)$ & $\pm52.3\,(2.3\sigma)$ & $\pm54.4\,(2.2\sigma)$ & $11.5\pm40.4\,(0.3\sigma)$ & $\pm23.9\,(0.5\sigma)$ & $\pm25.8\,(0.4\sigma)$ \\
 & & 500 & $68.0\pm54.6\,(1.2\sigma)$ & $\pm33.0\,(2.1\sigma)$ & $\pm30.1\,(2.3\sigma)$ & $16.5\pm23.8\,(0.7\sigma)$ & $\pm14.8\,(1.1\sigma)$ & $\pm15.1\,(1.1\sigma)$ \\ 
 & & 800 & $16.4\pm47.5\,(0.3\sigma)$ & $\pm24.8\,(0.7\sigma)$ & $\pm19.4\,(0.8\sigma)$ & $11.2\pm15.0\,(0.7\sigma)$ & $\pm11.8\,(0.9\sigma)$ & $\pm11.4\,(1.0\sigma)$
\enddata
\tablecomments{}
\label{nsig1}
\end{deluxetable*}

\begin{deluxetable*}{ccrrrrrrr}
\rotate
\tablenum{7} \tablecolumns{9} \tablecaption{Coherence Signals (Neighbors Divided by Color)} \tablewidth{0pt}
\tablehead{ & & & \multicolumn{3}{c}{Red Neighbors} & \multicolumn{3}{c}{Blue Neighbors} \\
Weighting & Angular & $D$ & $\langle\Delta v\rangle^{c}_{R-L} \pm\sigma_{\textrm{\tiny BST}}$ & $\pm\sigma_{\textrm{\tiny RAX}}$ & $\pm\sigma_{\textrm{\tiny RFA}}$ & $\langle\Delta v\rangle^{c}_{R-L} \pm\sigma_{\textrm{\tiny BST}}$ & $\pm\sigma_{\textrm{\tiny RAX}}$ & $\pm\sigma_{\textrm{\tiny RFA}}$\\
& momentum & [kpc] & [km s$^{-1}$] & [km s$^{-1}$] & [km s$^{-1}$] & [km s$^{-1}$] & [km s$^{-1}$] & [km s$^{-1}$]}
\startdata
Rel-L & $R\leq R_e$ & 30 & $101.8\pm143.5\,(0.7\sigma)$ & $\pm90.8\,(1.1\sigma)$ & $\pm104.6\,(1.0\sigma)$ & $53.5\pm92.9\,(0.6\sigma)$ & $\pm48.4\,(1.1\sigma)$ & $\pm61.8\,(0.9\sigma)$ \\
 & & 200 & $92.2\pm67.2\,(1.4\sigma)$ & $\pm58.0\,(1.6\sigma)$ & $\pm45.2\,(2.0\sigma)$ & $-3.7\pm58.1\,(0.1\sigma)$ & $\pm26.8\,(0.1\sigma)$ & $\pm36.5\,(0.1\sigma)$ \\
 & & 500 & $55.0\pm39.7\,(1.4\sigma)$ & $\pm30.9\,(1.8\sigma)$ & $\pm29.8\,(1.8\sigma)$ & $-20.2\pm29.1\,(0.7\sigma)$ & $\pm23.7\,(0.9\sigma)$ & $\pm25.1\,(0.8\sigma)$ \\
 & & 800 & $39.6\pm28.0\,(1.4\sigma)$ & $\pm26.5\,(1.5\sigma)$ & $\pm20.7\,(1.9\sigma)$ & $18.4\pm19.4\,(1.0\sigma)$ & $\pm22.7\,(0.8\sigma)$ & $\pm17.9\,(1.0\sigma)$ \\ 
\hline
Rel-L & $R_e<R\leq 2R_e$ & 30 & $161.9\pm81.8\,(2.0\sigma)$ & $\pm87.4\,(1.9\sigma)$ & $\pm63.9\,(2.5\sigma)$ & $66.7\pm38.5\,(1.7\sigma)$ & $\pm65.5\,(1.0\sigma)$ & $\pm79.4\,(0.8\sigma)$ \\  
 & & 200 & $124.2\pm69.9\,(1.8\sigma)$ & $\pm63.6\,(2.0\sigma)$ & $\pm45.0\,(2.8\sigma)$ & $80.2\pm77.5\,(1.0\sigma)$ & $\pm82.3\,(1.0\sigma)$ & $\pm42.4\,(1.9\sigma)$ \\ 
 & & 500 & $75.8\pm43.3\,(1.7\sigma)$ & $\pm34.5\,(2.2\sigma)$ & $\pm25.4\,(3.0\sigma)$ & $36.0\pm34.2\,(1.1\sigma)$ & $\pm43.1\,(0.8\sigma)$ & $\pm30.8\,(1.2\sigma)$ \\
 & & 800 & $64.9\pm29.3\,(2.2\sigma)$ & $\pm26.9\,(2.4\sigma)$ & $\pm16.4\,(4.0\sigma)$ & $57.1\pm20.9\,(2.7\sigma)$ & $\pm31.9\,(1.8\sigma)$ & $\pm21.7\,(2.6\sigma)$ \\ 
\hline
Abs-L & $R\leq R_e$ & 30 & $77.3\pm163.3\,(0.5\sigma)$ & $\pm102.8\,(0.8\sigma)$ & $\pm100.7\,(0.8\sigma)$ & $-5.6\pm111.2\,(0.1\sigma)$ & $\pm51.9\,(0.1\sigma)$ & $\pm64.5\,(0.1\sigma)$ \\
 & & 200 & $73.0\pm54.3\,(1.3\sigma)$ & $\pm34.7\,(2.1\sigma)$ & $\pm31.4\,(2.3\sigma)$ & $9.4\pm44.7\,(0.2\sigma)$ & $\pm31.0\,(0.3\sigma)$ & $\pm37.8\,(0.2\sigma)$ \\  
 & & 500 & $40.5\pm31.8\,(1.3\sigma)$ & $\pm19.6\,(2.1\sigma)$ & $\pm18.4\,(2.2\sigma)$ & $1.5\pm25.7\,(0.1\sigma)$ & $\pm28.9\,(0.1\sigma)$ & $\pm35.6\,(0.0\sigma)$ \\  
 & & 800 & $12.6\pm22.9\,(0.5\sigma)$ & $\pm15.2\,(0.8\sigma)$ & $\pm13.6\,(0.9\sigma)$ & $-1.6\pm15.8\,(0.1\sigma)$ & $\pm21.4\,(0.1\sigma)$ & $\pm23.6\,(0.1\sigma)$ \\
\hline
Abs-L & $R_e<R\leq 2R_e$ & 30 & $164.6\pm94.0\,(1.8\sigma)$ & $\pm99.6\,(1.7\sigma)$ & $\pm131.7\,(1.2\sigma)$ & $-4.4\pm80.4\,(0.1\sigma)$ & $\pm68.3\,(0.1\sigma)$ & $\pm88.4\,(0.0\sigma)$ \\ 
 & & 200 & $86.3\pm55.1\,(1.6\sigma)$ & $\pm35.9\,(2.4\sigma)$ & $\pm37.7\,(2.3\sigma)$ & $17.3\pm48.5\,(0.4\sigma)$ & $\pm35.0\,(0.5\sigma)$ & $\pm39.6\,(0.4\sigma)$ \\ 
 & & 500 & $50.2\pm33.0\,(1.5\sigma)$ & $\pm20.2\,(2.5\sigma)$ & $\pm17.3\,(2.9\sigma)$ & $6.2\pm27.2\,(0.2\sigma)$ & $\pm30.6\,(0.2\sigma)$ & $\pm39.5\,(0.2\sigma)$ \\  
 & & 800 & $17.6\pm23.8\,(0.7\sigma)$ & $\pm15.9\,(1.1\sigma)$ & $\pm12.6\,(1.4\sigma)$ & $3.4\pm18.0\,(0.2\sigma)$ & $\pm23.0\,(0.1\sigma)$ & $\pm26.4\,(0.1\sigma)$
\enddata
\tablecomments{}
\label{nsig2}
\end{deluxetable*}

\begin{deluxetable}{cccccc}
\tablenum{8} \tablecolumns{6} \tablecaption{Coherence Signals (Summary)} \tablewidth{0pt}
\tablehead{  & & \multicolumn{2}{c}{Rel-L} & \multicolumn{2}{c}{Abs-L} \\
Sample & Rotation & $D1$ & $D2$ & $D1$ & $D2$ }
\startdata
Whole & Central &  - - - & $\ast$ - $\ast$ & - $\ast\ast$ & - - - \\
& Outskirt & $\circ\ast\bullet$ & $\bullet\bullet\bullet$ & - $\circ\ast$ & - $\ast\ast$ \\
\hline \hline
Bright & Central &  - - - & - - - &  - - - & - - - \\
& Outskirt & - - - & - - - &  - - - & - - - \\
\hline
Faint & Central & - - - & $\ast$ - $\ast$ & - $\ast\circ$ & - - - \\
& Outskirt & $\circ\ast\bullet$ & $\bullet\bullet\bullet$ & $\ast\circ\bullet$ & $\circ\ast\circ$ \\
\hline \hline
Red & Central & - - - & - - - & - - - & - - - \\
& Outskirt & - - $\ast$ & - - -  & - - - & - - - \\
\hline
Blue & Central & - - - & - - - & - - $\ast$ & - - - \\
& Outskirt & $\ast$ - $\bullet$ & $\bullet\circ\bullet$ & $\ast$ - $\ast$ & $\ast\ast\circ$ \\
\hline \hline
Concentrated & Central & - - $\bullet$ & - - $\ast$ & - - - & - - - \\
& Outskirt & - - $\bullet$ & $\ast\ast\bullet$ & - - - & - - - \\
\hline
Diffuse & Central & - - - & - - - & - - - & - - - \\
& Outskirt & $\bullet$ - $\circ$ & $\circ\ast\circ$ & $\bullet$ - $\ast$ & - - $\ast$ \\
\hline \hline
Well-aligned & Central & - - - & - - - & - - - & - - - \\
& Outskirt & - - - & - - - & - - - & - - - \\
\hline
Misaligned & Central & - - - & $\ast$ - - & - - - & $\ast\ast\ast$ \\
& Outskirt & $\circ\ast\bullet$ & $\circ\circ\bullet$ & - $\circ\ast$ & $\circ\circ\circ$ \\
\hline \hline
Bright & Central &  - - - & - - $\ast$ &  - - - & - - - \\
Neighbors & Outskirt & $\circ\ast\bullet$ & $\ast\bullet\bullet$ & $\ast\ast\ast$ & - $\ast\ast$ \\
\hline
Faint & Central &  - - - & - - - &  - - - & - - - \\
Neighbors & Outskirt &  - - - & - - - &  - - - & - - - \\
\hline \hline
Red & Central & - - $\ast$ & - - - & - $\ast\ast$ & - $\ast\ast$ \\
Neighbors & Outskirt & $\ast\ast\circ$ & $\ast\ast\bullet$  & - $\ast\ast$ & - $\circ\circ$ \\
\hline
Blue & Central & - - - & - - - & - - - & - - - \\
Neighbors & Outskirt & - - - & $\circ$ - $\circ$  & - - - & - - -
\enddata
\tablecomments{In each table cell, the ratios of $\langle\Delta v\rangle^{c}_{R-L}$ signal to BST, RAX, and RFA uncertainties are simplified as follows: $\bullet$ ($\ge 3.0\sigma$); $\circ$ ($2.5 - 2.9\sigma$); $\ast$ ($2.0 - 2.4\sigma$); and - ($\le 1.9\sigma$). $D1$: $D=30$ or 200 kpc. $D2$: $D=500$ or 800 kpc. Several cases with negative signals (but in $<3\sigma$) are denoted by `-'.}
\label{signalsum}
\end{deluxetable}

\acknowledgments
This study uses data provided by the Calar Alto Legacy Integral Field Area (CALIFA) survey (http://califa.caha.es/), which is based on observations collected at the Centro Astron{\'o}mico Hispano Alem{\'a}n (CAHA) at Calar Alto, operated jointly by the Max-Planck-Institut f{\"u}r Astronomie and the Instituto de Astrof{\'i}sica de Andaluc{\'i}a (CSIC).
This study also uses the the NASA/IPAC Extragalactic Database (NED), which is operated by the Jet Propulsion Laboratory, California Institute of Technology, under contract with the National Aeronautics and Space Administration.

\end{document}